%% file: main.tex
\documentclass[12pt]{article}

\usepackage{microtype}
\usepackage{amsthm}
\usepackage{graphicx}
\usepackage{fontawesome}
\usepackage{multirow}
\usepackage{float}
\usepackage{acronym}
\usepackage{listings}
\usepackage{subcaption}
\usepackage{booktabs}
\usepackage{array}
\usepackage{todonotes}
\usepackage{svg}
\usepackage{tikz}
\usetikzlibrary{cd}
\usetikzlibrary{matrix,calc}
\usepackage{url}
\usepackage[breaklinks]{hyperref}
\usepackage{amsmath}
\usepackage{amssymb}
\usepackage{mathtools}
\usepackage{amsthm}
\usepackage{bm}
\usepackage[round]{natbib}
\usepackage{amsmath, amssymb, graphicx, url, mathtools}
\usepackage[linesnumbered,ruled,vlined]{algorithm2e} 

\usepackage{subcaption}
\usepackage{wrapfig}
\usepackage{tikz}
\usepackage{multirow}
\usepackage{mathptmx}

\newtheorem{prop}{Proposition}
\newtheorem{assumption}{Assumption}
\newtheorem{cor}{Corollary}
\theoremstyle{remark}
\newtheorem{remark}{Remark}

\newcommand{\blind}{1}
\begin{document}

\def\spacingset#1{\renewcommand{\baselinestretch}%
{#1}\small\normalsize} \spacingset{1}


\if1\blind
{
  \title{\bf Adventures in Demand Analysis Using AI}
  \author{
    Philipp Bach\thanks{Freie Universität Berlin, \texttt{philipp.bach@fu-berlin.de}}\hspace{.2cm} \\
    Victor Chernozhukov\thanks{Massachusetts Institute of Technology, \texttt{vchern@mit.edu}} \\
    Sven Klaassen\thanks{University of Hamburg, \texttt{svenklaassen92@gmail.com}} \\
    Martin Spindler\thanks{University of Hamburg, \texttt{martin.spindler@uni-hamburg.de}} \\
    Jan Teichert-Kluge\thanks{University of Hamburg, \texttt{jan.teichertkluge@uni-hamburg.de}} \\
    and \\
    Suhas Vijaykumar\thanks{UC San Diego, \texttt{svijaykumar@ucsd.edu}}
  }
  \maketitle
} \fi

\if0\blind
{
  \bigskip
  \bigskip
  \bigskip
  \begin{center}
    {\LARGE\bf Adventures in Demand Analysis Using AI}
  \end{center}
  \medskip
} \fi

\bigskip
\begin{abstract}

\noindent This paper advances empirical demand analysis by integrating multimodal product representations derived from artificial intelligence (AI). 
Using a detailed dataset of toy cars on \textit{Amazon.com}, we combine text descriptions, images, and tabular covariates to represent each product using transformer-based embedding models. 
These embeddings capture nuanced attributes, such as quality, branding, and visual characteristics, that traditional methods often struggle to summarize. 
Moreover, we fine-tune these embeddings for causal inference tasks. 
We show that the resulting embeddings substantially improve the predictive accuracy of sales ranks and prices and that they lead to more credible causal estimates of price elasticity. 
Notably, we uncover strong heterogeneity in price elasticity driven by these product-specific features. 
Our findings illustrate that AI-driven representations can enrich and modernize empirical demand analysis. 
The insights generated may also prove valuable for applied causal inference more broadly.

\smallskip

\noindent%
\vfill

\end{abstract}

\newpage
\spacingset{1.75} 

\input{JASA_Buybox_Version/Introduction}

\input{JASA_Buybox_Version/Section2}
\input{JASA_Buybox_Version/Section3}

\input{JASA_Buybox_Version/Conclusion}

\appendix
\input{JASA_Buybox_Version/Section4}

\bibliographystyle{abbrvnat}
\bibliography{bibliography-clean-short}

\clearpage 
\input{JASA_Buybox_Version/OnlineAppendix}

\end{document}

%% file: JASA_Buybox_Version/Introduction.tex
\section{Introduction}
Almost a century ago, \textit{The Journal of the American Statistical Association} published early empirical studies of demand that applied statistical methods to measure how consumers respond to price changes. Work by scholars such as 
\citet{Wright1929Review}, 
\citet{working1943statistical}, 
\citet{schultz1933standard},
\citet{mills1931use, Mills1937,mills1937movements}, and
\citet{stigler1939limitations} 
moved economics from theory towards quantitative measurement. By doing so, this work provided a foundation for econometrics, a field of statistical analysis focusing on economic problems. Their research established a tradition of using data to understand market behavior and inform economic models.

Today, advances in artificial intelligence (AI) and machine learning offer opportunities to build on this tradition. Instead of relying solely on simple numeric variables, researchers can now incorporate AI-generated product representations derived from text descriptions and images. These methods draw on hedonic modeling approaches \citep{griliches1971hedonic, pakes2003reconsideration} and integrate recent machine learning techniques \citep{devlin2019bert, dosovitskiy2021image}, allowing economists to represent products more richly and capture nuances that standard covariates do not.

Using sales ranking and price data for toy cars on \textit{Amazon.com}, we demonstrate how transformer-based models can leverage multiple, rich sources of product information for demand analysis. Our data include text descriptions, images, sales ranks, and prices.
These multimodal inputs yield highly informative numerical embeddings that capture demand-relevant product attributes not easily summarized by standard human-encoded tabular variables---such as quality, branding, and visual characteristics---as we illustrate in Section \ref{sec-repr-products}.

We then fine-tune these embeddings to predict price and quantity signals, as these predictions are critical inputs to our causal inference problem.  The resulting models achieve higher predictive accuracy than simpler specifications which rely solely on tabular data. 
Embeddings capture subtle distinctions between products—such as quality, branding, or visual characteristics—that influence consumer demand and market prices, but are difficult to quantify using conventional methods. 
This improvement in predictive power suggests that AI-generated representations can meaningfully enhance empirical demand analysis and other causal inference tasks.

Finally, we address the challenge of estimating the price elasticity of demand, a central economic parameter. In our setting, simple cross-sectional regressions yield implausibly small elasticity estimates because they fail to capture product visibility and quality as key confounders. This motivates us to formulate a dynamic model with multimodal product attributes, along with lagged quantity and price signals, all of which serve both as confounders and as price-elasticity modifiers. By estimating such a dynamic model, we obtain more realistic price elasticities. Furthermore, we uncover pronounced heterogeneity in price elasticities that varies with product characteristics, as well as with how expensive and popular the products are. This underscores the economic value of AI-based representations: when properly fine-tuned, they yield more nuanced and credible estimates of how consumers respond to price changes across different products.

Our approach contributes to multiple strands of the literature. It extends empirical demand analysis by employing AI-generated, multimodal representations of products. Our work also builds on the emerging intersection of econometrics and machine learning \citep{belloni2014high,varian2014big, athey2019machine, mullainathan2017machine, chernozhukov2018} and complements recent studies that apply AI-based text analysis and other modern methods to economic questions \citep{bajari2023hedonic, compiani2023demand}. In doing so, it provides a framework for combining flexible product representations with established econometric tools for identification and inference. It also introduces the idea that embeddings can and should be fine-tuned with causal inference in mind---which is critical in our context and potentially useful in other applications.

Our main empirical result is that AI-based embeddings are important effect \textit{modifiers}: they capture substantial heterogeneity in price elasticity across products. Interestingly, our findings also suggest that these embeddings are not major confounders of the price-quantity relationship. Although highly predictive of price and quantity levels, they are weak predictors of the temporal changes we use to identify elasticities. Thus, while standard homogeneous price-effect models can yield reasonable estimates of the average elasticity, they can severely understate or overstate elasticity for certain sets of products. These findings constitute important new empirical insights.

Our work also adds to a nascent literature in applied industrial organization that estimates demand models using public e-commerce data—such as ratings, reviews, or sales rankings—as proxies for the quantity sold, as the quantity sold is generally not published \citep{Waldfogel2021,he2020sales,Musolff2021}. Building on \citet{he2020sales} (see also \citealp{goolsbee2002measuring}), we treat the sales ranking as a proxy for relative quantity sold, motivated by the relationship among order statistics of the power law distribution. We show that combining this approach with high-quality embeddings of product descriptions and images produces realistic estimates of price elasticities.

The remainder of the paper is organized as follows: Section \ref{sec-repr-products} discusses the use of AI-driven representations in demand analysis and describes the toy cars dataset, including how we extract and process multimodal features to create product embeddings. It also presents our first empirical results, highlighting how embeddings improve accuracy in predicting prices and quantities. Section \ref{sec:sensitivity} focuses on estimating the price elasticity; it lays out the underlying causal inference problem, discussing potential sources of confounding. We also illustrate the power of our embeddings in describing the heterogeneity of the price elasticity function.  Finally, Section \ref{sec:conclusion} concludes by summarizing our findings and discussing their implications for future research at the intersection of AI and econometrics. The Addendum contains deferred theoretical discussions, and the Online Appendix describes the workflow and algorithms used for feature generation and model estimation.

%% file: JASA_Buybox_Version/Section2.tex
\section{Using AI to Understand and Represent Products} \label{sec-repr-products}

\subsection{The Data and Measurement of Prices and Quantities }\label{sec:data}

\begin{figure}[h]
    \centering
    \includegraphics[width=0.5\textwidth]{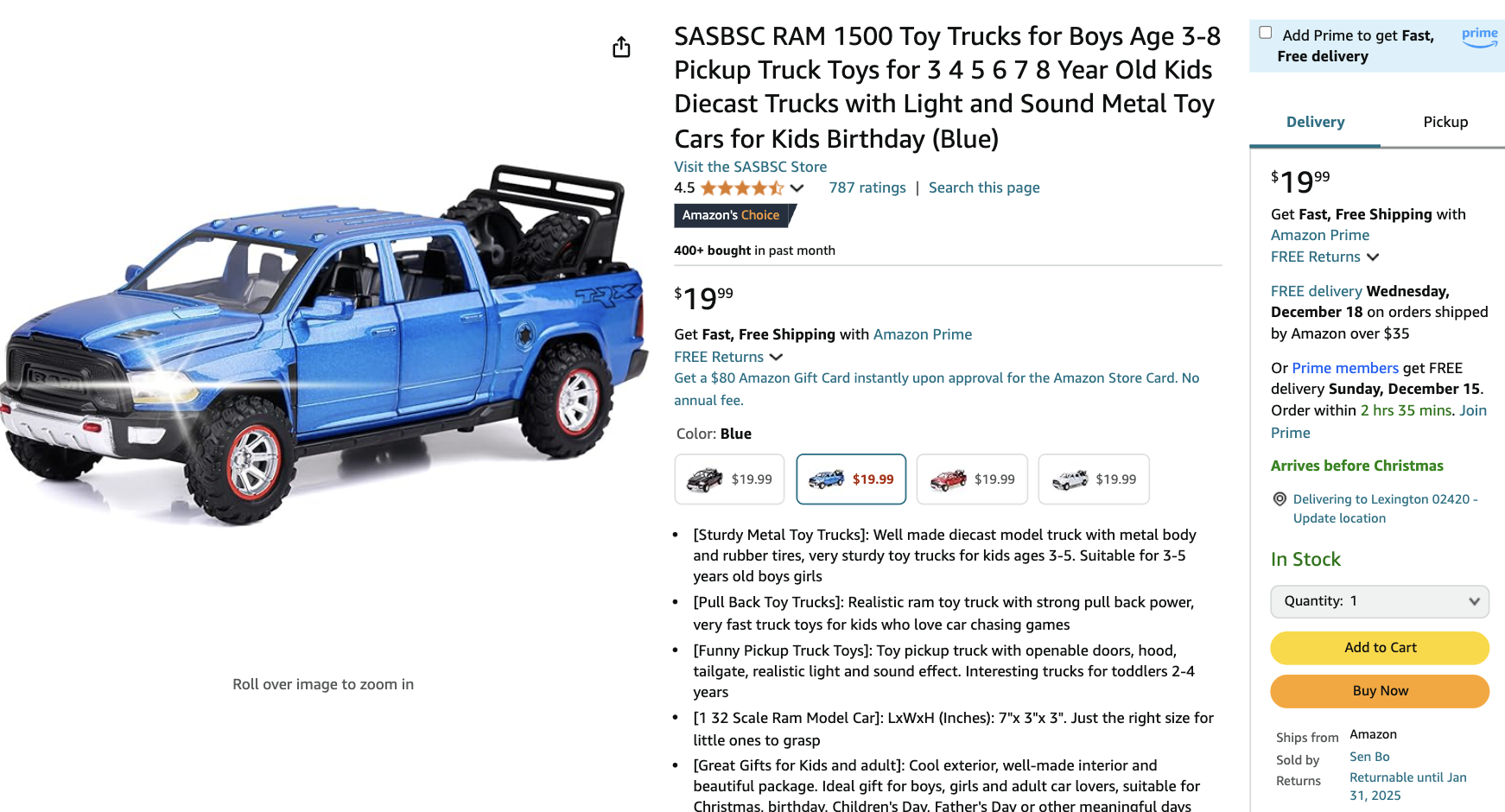}
    \caption{A product example with an image and text description in the "toys" category.}
    \label{fig:toyexample}
\end{figure}

Our analysis uses a data set of toy cars from \textit{Amazon.com}, compiled and provided by the data aggregator \textit{Keepa.com}. For each item \(i\), we collected its sales rank and price at time points spanning from March 2023 to January 2024. We also gathered each product's description, image, and additional tabular features (e.g., its subcategory on \textit{Amazon.com}), as summarized in Table \ref{tab:variables}. Figure \ref{fig:toyexample} illustrates a typical product page containing the product image and description. Overall, our data set comprises \(N = 7{,}226\) unique products.

For our analysis, we define the quantity signal as
\[
Q_{it} = \log\!\bigl(1/\text{Time-Averaged Sales Rank of } i \text{ in period } t \bigr),
\]
and the price signal as
\[
P_{it} = \log\!\bigl(\text{Time-Averaged Price of } i \text{ in period } t\bigr).
\]
Each period, indexed by \(t = 1, \ldots, T\), spans 4 weeks. We have \(T = 12\) periods in total, which are directly adjacent to each other. We structure our data set this way to limit inter-temporal feedback in our price elasticity analysis in Section \ref{sec:sensitivity}. We also examine temporal changes in these signals, $
\Delta Q_{it} := Q_{it} - Q_{i(t-1)}$ and $
\Delta P_{it} := P_{it} - P_{i(t-1)}.$
\begin{figure}[ht]
    \centering
     \includegraphics[height=0.3\textwidth]{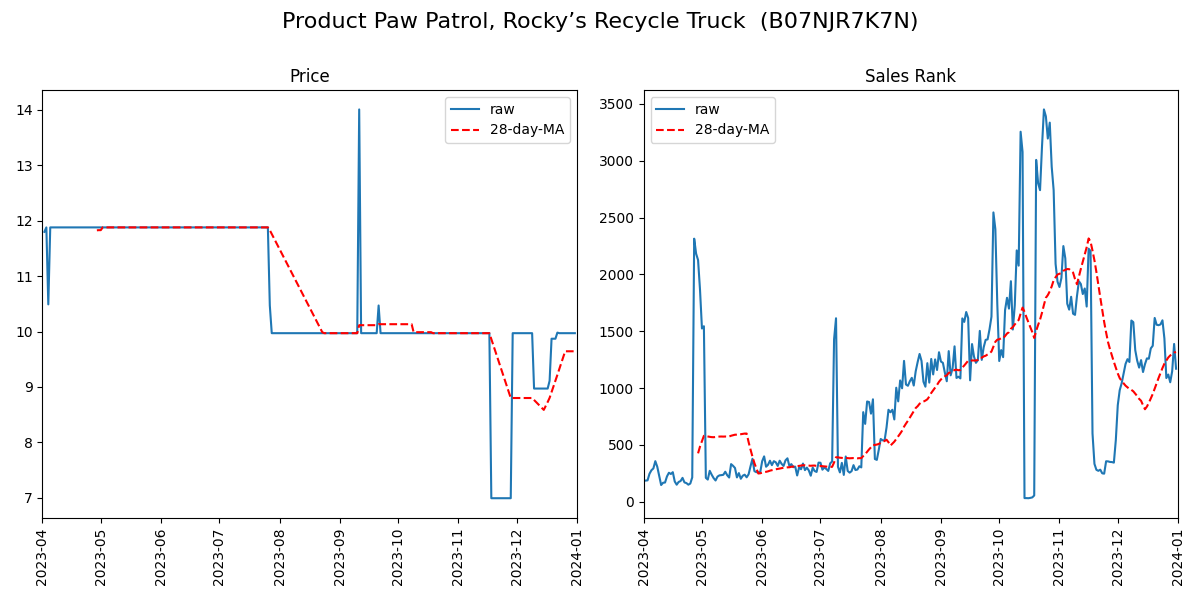}\\
    \includegraphics[height=0.3\textwidth]{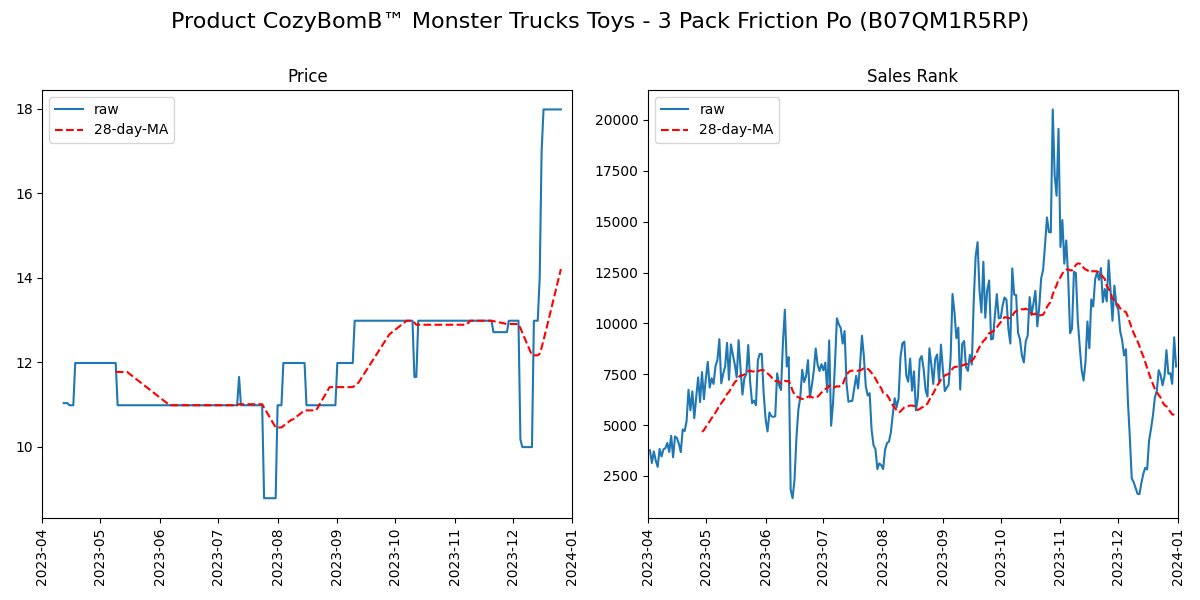} 
    \caption{Price and sales rank series  for two example products.}
    \caption*{\footnotesize Note: The main series are shown as solid lines. The 28-day moving averages are shown in dashes.}
    \label{fig:example}
\end{figure}

In what follows, we refer to \(Q_{it}\) and \(P_{it}\) as the quantity and price signals, respectively. Under a power law assumption (see Remark \ref{rmk:time-avg} below), the logarithm of actual sales is proportional to the logarithm of the inverse sales rank; thus, we adopt inverse sales rank as our quantity signal. Furthermore, under this assumption, the price sensitivity of the inverse rank is proportional to the price sensitivity of the actual quantity sold, enabling us to capture demand responses to price changes.

\begin{table}[ht]\footnotesize
    \centering
    \caption{Variable description, toy products data set.}
    \begin{tabular}{lp{12cm}}
        \hline \hline
         \textbf{Variable} & \textbf{Description}  \\ \hline
         Sales rank & Ranking by units sold, relative to other products in the ``games and toys'' category \\
         Price & Buy Box price for a new product, excluding shipping and handling fees \\
         Review count  & Number of reviews  \\
         Rating  & Average customer feedback rating \\
         Lightning deals & Binary; 1 if product is promoted via a lightning deal \\
         Buy box is FBA & Binary; 1 if the product delivery is fulfilled by Amazon \\
         Product description  & Unstructured text; includes title, manufacturer, brand, model, color, and size \\
         Subcategory & Categorical; product subcategory \\
         Product image  & First image featured on the product page \\
         ASIN & Unique product ID \\ \hline \hline
    \end{tabular}
    \caption*{\footnotesize Note: Product description, image, product ID, and subcategory do not vary with time in our data set; all other variables can vary with time.}
    \label{tab:variables}
\end{table}

\begin{remark}\label{rmk:time-avg}
Our use of the time-averaged inverse sales rank is an approximation motivated by modeling the latent, true quantities \(Q^*_{it}\) as independent draws from an underlying distribution at each time \(t\).
If an item has sales rank \(k\) at time \(t\), the quantity sold should be distributed as the \((n-k+1)^{\text{th}}\) order statistic from a sample of size \(n\), denoted $Q_{(n-k+1)t}^*$. In the case of Pareto distributions with shape \(\vartheta\), i.e. $\Pr(Q^*_{it} > t) = C_\vartheta\cdot t^{-\vartheta}$, we have the asymptotic approximation \(\log\mathrm{E}[Q_{(n-k+1)t}] \sim C_{\vartheta,n} -\vartheta^{-1}\log(k).\)\footnote{Moments of order statistics for the Pareto distribution are given explicitly by \citet{malik1966exact} (see also \citealp{vannman1976estimators}); combining the result with Stirling's approximation gives $\mathrm{E}[Q_{(n-k+1)t}^*] \sim C_\vartheta(k/n)^{-1/\vartheta}$ as $n \to \infty$.} This motivates our use of the negative log sales rank, $Q_{it}$. Using this approximation, \citet{he2020sales} estimate \(\hat\vartheta \approx 0.5\) for toys on \textit{Amazon.com}; consequently, we can multiply our estimates by \(1/\hat\vartheta \approx 2\) to obtain rough estimates of the price elasticity of demand. More generally, one can quantify the connection between our estimates and price elasticities under various assumptions about the distribution of sales \citep{office2020using}.
\end{remark}
\subsection{Using AI to Represent Products}

To convert product data into useful numerical features, we employ various encoding models based upon the transformer neural network architecture proposed by \citet{vaswani2017attention}. We convert text descriptions into  embeddings \(T_{i}\) using language models such as RoBERTa \citep{liu2019robertarobustlyoptimizedbert} or LLaMA \citep{touvron2023llamaopenefficientfoundation}, convert images into dense embeddings \(I_{i}\) using the BEiT model \citep{bao2022beit}, and transform tabular data into embeddings using the SAINT model \citep{somepalli2021saintimprovedneuralnetworks}. The Online Appendix provides additional implementation details.

While the aforementioned models are designed to work with the data we have on hand, the underlying approach is well suited for generalization to new types of data. 
In order to succeed in our context, these models must also be integrated and fine-tuned appropriately for estimating the price sensitivity. 
We discuss three key aspects of this approach which help explain its success and facilitate generalization to new contexts: self-supervised learning, the attention-based transformer architecture, and fine-tuning motivated by orthogonalized estimation of causal effects.

\noindent\textbf{Self-Supervision.}  
A significant challenge in machine learning is the scarcity of high-quality labeled data, as manual annotation is both expensive and time-consuming. Self-supervision addresses this limitation by creating labeled examples directly from unlabeled data. In this process, a portion of the input is deliberately masked or corrupted, and the model learns to predict these masked elements. This approach, fundamental to models like BERT \citep{devlin2019bert}, effectively transforms each input sample into a self-labeled instance.

Consider the example sentence:
$
S = \text{``Well made diecast model truck with metal body.''}
$
We create a masked version:
$W = \text{``Well made [m] model truck with [m] body''}.$
The original sequence \(S\) serves as an auxiliary label that the model attempts to reconstruct from the corrupted input \(W\). By applying this approach to billions of sentences, models learn to capture syntactic and semantic relationships without explicitly annotated labels.

The resulting internal representations, called \emph{embeddings}, are extracted from the model's hidden layers and represent features of words or sentences. The approach generalizes well to other data types: for images, masking out patches and asking the model to predict the missing parts \citep{he2022masked} enables the extraction of informative, context-dependent embeddings.

\noindent\textbf{Attention.}  
Transformer-based models employ so-called \emph{attention mechanisms} \citep{vaswani2017attention} to efficiently represent data; these underlie all of the embedding models used in this paper, as well as the highly influential GPT models \citep{mann2020language}. Attention refers to a specific  structure repeated several times within transformer neural networks, which allows the model to selectively weight the most relevant components of the input when making predictions. 

Through adaptive weighting of different input elements, attention produces embeddings that incorporate contextual and nuanced relationships between objects. This is believed to produce more informative embeddings than earlier, context-free models (such as  word2vec or GloVe in the case of word embeddings;  \citealp{mikolov2013efficient, pennington2014glove}). Quantitatively, attention-based models achieve state-of-the-art performance across a wide range of tasks \citep{mann2020language,bao2022beit}.

\noindent\textbf{Causal Fine-Tuning.}  
After a model has learned embeddings through self-supervision, it can be adapted for various downstream tasks. In our setting, the embeddings are important inputs to our causal inference problem: we are interested in how prices affect demand, holding fixed both product characteristics and other demand determinants.  Our \emph{fine-tuning} updates the pre-trained model parameters for this specific end-goal, by optimizing prediction of quantity and price signals. This is precisely the right target for orthogonal estimation of the price elasticity, as we further discuss in Section \ref{subsec:orthogonal-inference}.

During fine-tuning, the embeddings serve as inputs to a specialized prediction layer. The errors from the prediction
layers are then used to inform and update the parameters of the embeddings through gradient descent steps, which are computed via back-propagation \citep{rumelhart1986learning}.  
The following diagram summarizes the process: 
\begin{center}
\begin{tikzcd}
[/tikz/column 2/.append style={anchor=base east}]
 & \quad A^{tx}_{i} &
\\
X^{in}_{i} =  \begin{bmatrix}
    \mathrm{Text}_{i} \\
    \mathrm{Image}_{i}
\end{bmatrix}
\arrow[r, maps to, "e"] & E_i \coloneqq \begin{bmatrix} T_{i} \\ I_{i} \end{bmatrix} \arrow[u, maps to, shift right=2] \arrow[d, maps to, shift left=2] \arrow[r, maps to, "m"] & \{\hat Q_{it}, \hat P_{it}\}_{t=1}^T. \\
 & \quad A^{im}_{i} &
\end{tikzcd}
\end{center}
Here, \(X^{in}_{i}\) represents the raw inputs (text \(\mathrm{Text}_{i}\) and image \(\mathrm{Image}_{i}\)). The embedding map \(e\) transforms these inputs into embeddings~\(E_{i}\). The terms \(A^{im}_{{i}}\) and \(A^{tx}_{{i}}\) are the auxiliary (masked) targets or ``labels'' in the self-supervised task. The model learns embeddings by attempting to reconstruct these targets from the unmasked parts of the inputs. The map \(m\) represents a downstream prediction layer that uses the embeddings \(E_{i}\) to predict tasks of interest, such as \(\hat Q_{it}\) (quantities) and \(\hat P_{it}\) (prices) over time.

As the model learns to reconstruct the auxiliary targets, it refines its internal representations. These improved embeddings \(E_{i}\) are then used by \(m\) to make high-quality predictions. Fine-tuning adjusts both \(e\) and \(m\) to ensure that the embeddings and downstream predictions align with the target predictive and causal inference questions.

\subsection{Evaluating the Embeddings}
After obtaining the embeddings, we must assess whether they effectively represent the products and “understand” their characteristics. We first take the concatenated embeddings \(E_i =\bigl (T_{i}, I_{i}\bigr)\), where \(T_{i}\) also includes tabular embeddings, and then apply a Johnson--Lindenstrauss projection of these embeddings onto a 256-dimensional vector \(\,\bar E_{i}\); \cite{johnson1984extensions}. This projection approximately preserves distances and is therefore considered (at least approximately) information-preserving. We then center and normalize the embeddings so they lie on a hypersphere:
\[
X^e_{i} := \frac{\bar E_{i} - \frac{1}{n} \sum_i \bar E_{i}}{\bigl\|\,\bar E_{i} - \tfrac{1}{n} \sum_i \bar E_{i}\bigr\|}.
\]
We use these normalized embeddings in our subsequent analysis.

We evaluate these embeddings through two approaches:
\begin{itemize}
    \item[1.] \textbf{Qualitative.} We examine similar products or clusters of products on this hypersphere and assess the results qualitatively.
    \item[2.] \textbf{Quantitative.} We determine whether these AI-generated features improve predictions of price and quantity signals, where predictions serve as key inputs into downstream causal inference. 
\end{itemize}
Both approaches are crucial for demand analysis, including the computation of hedonic inflation prices, forecasting demand and prices for new products, and understanding how demand responds to price variations.

\subsubsection{Qualitative Assessment}

\begin{figure}[H]
    \centering
    \begin{minipage}{0.49\textwidth}
        \centering
        \includegraphics[width=\linewidth]{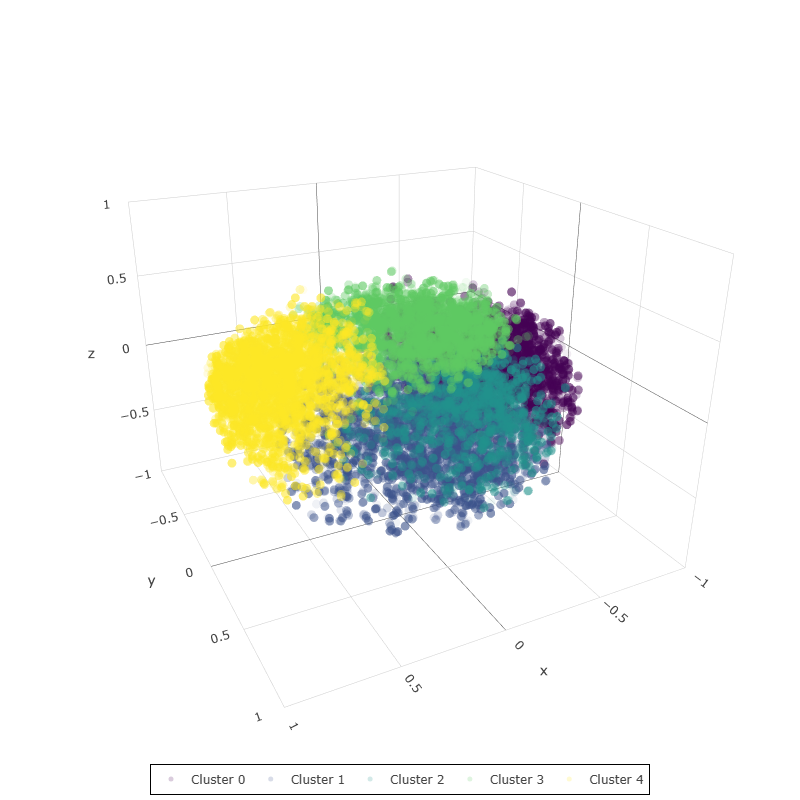}
        \caption{3D-representation of product embeddings (with image) and five clusters }
        \label{fig:cluster_pca_img_3d_comb}
    \end{minipage}
    \hfill
    \begin{minipage}{0.49\textwidth}
        \centering
        \includegraphics[width=\linewidth]{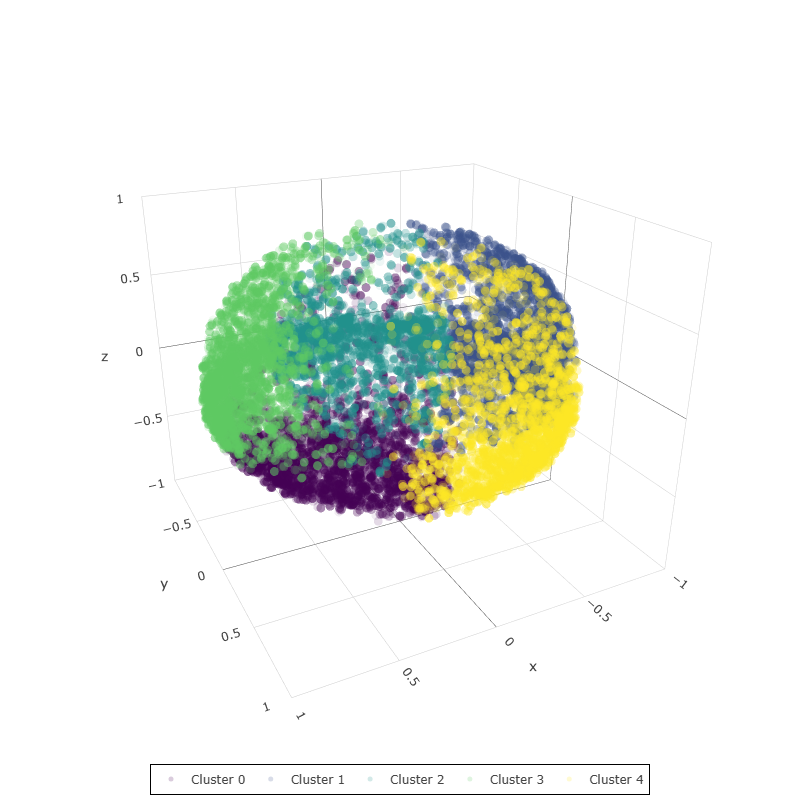}
        \caption{3D-representation of product embeddings (no image) and five clusters}
        \label{fig:cluster_pca_txt_3d}
    \end{minipage}
\end{figure}

For the clustering task, we perform $k$-means clustering to group products into five clusters based on their embeddings. To examine the influence of images, we first cluster using both text and image embeddings, and then using only text embeddings. We visualize the resulting product clusters in three-dimensional space by projecting the embeddings onto the first three principal components, as shown in Figures \ref{fig:cluster_pca_img_3d_comb} and \ref{fig:cluster_pca_txt_3d}.

When text and image embeddings are combined, the projection yields a “full” ball of product points with distinctly separated clusters. In contrast, using text-only embeddings produces a “stripe on a sphere,” where the points are concentrated near the boundary and around the equator of the ball. Nevertheless, the clusters remain well-separated even without the image information.

While mainly illustrative, these visuals suggest that text-only embeddings lie in a lower-dimensional space, missing valuable image-based information absent from text descriptions. This observation is supported by Tables \ref{fig:cluster_pca_img_3d_combined_table} and \ref{fig:cluster_pca_img_3d_comb}, which show that clusters formed from text+image embeddings have more visually coherent centroids and greater internal homogeneity—confirming the importance of multimodal data.

To further explore these clusters, we employ generative AI tools to summarize and characterize each cluster centroid. We also construct an “average” representative image for products near the centroids, with results shown in Table \ref{tab:gen_ai_summary}. These results closely match our own assessments of the product cluster centers, underscoring the utility of combining both text and image embeddings for cluster analysis.

\begin{table}[ht]
    \centering  \footnotesize
    \begin{tabular}{p{0.15\textwidth} c}
        \vspace{-1cm}{Cluster 0} & \includegraphics[height=.08\textheight]{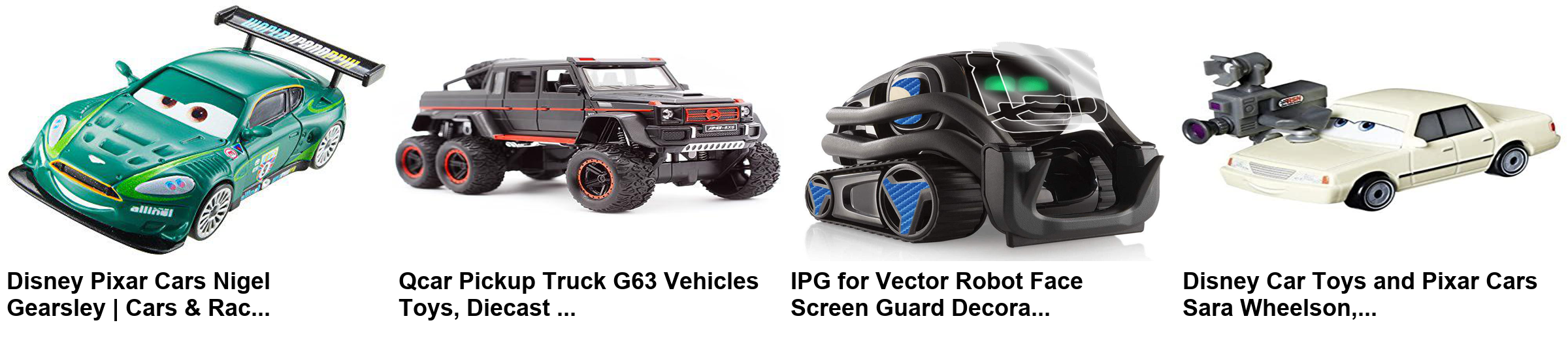} \\[0.1cm]
        \vspace{-1cm}{Cluster 2} & \includegraphics[height=.08\textheight]{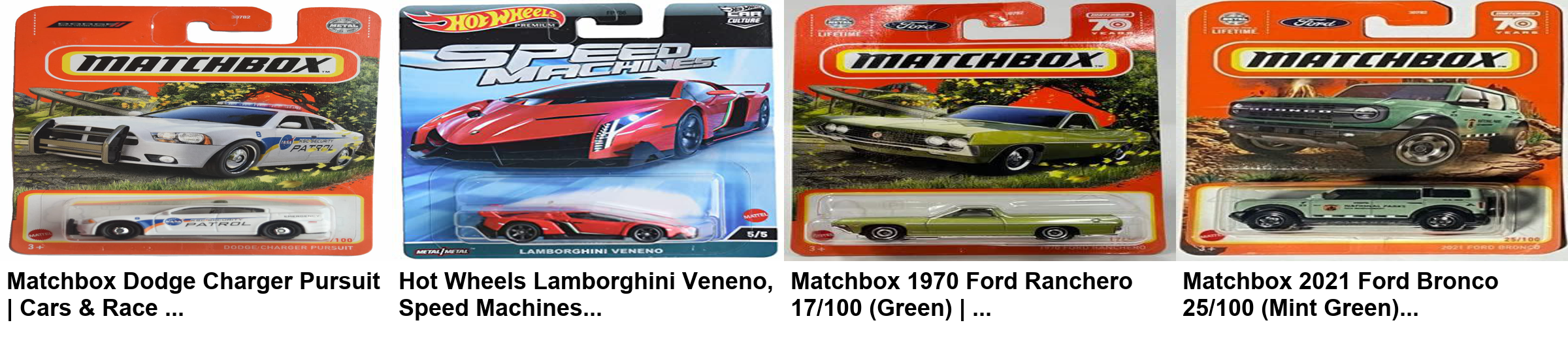} \\[0.1cm]
        \vspace{-1cm}{Cluster 4} & \includegraphics[height=.08\textheight]{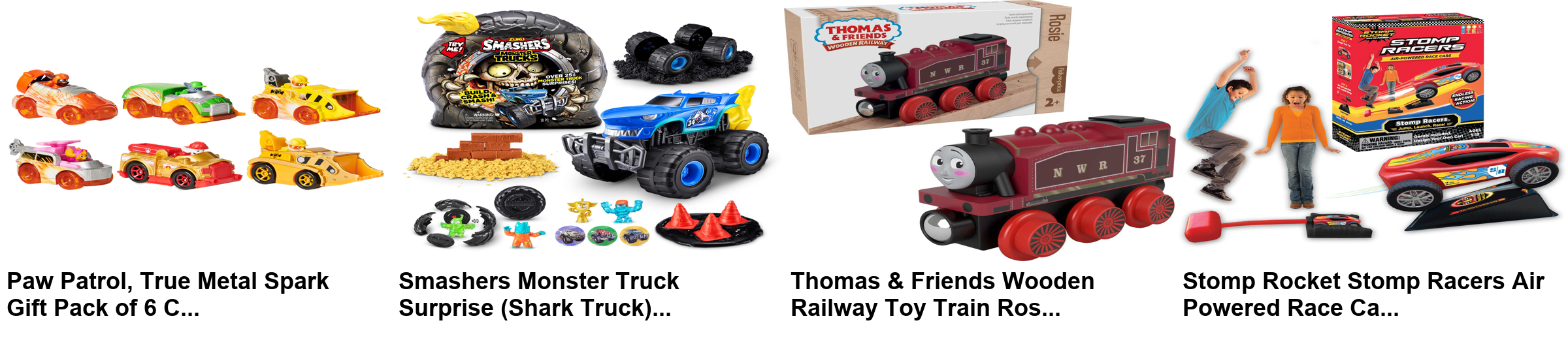} \\
    \end{tabular}
    \caption{Examples of closest products to cluster centers (Tabular + Text + Image); Only three examples of clusters shown. }
    \label{fig:cluster_pca_img_3d_combined_table}
\end{table}

\begin{table}[ht]  \footnotesize
    \centering
    \begin{tabular}{>{\centering\arraybackslash}p{0.15\textwidth} c}
        \vspace{-1cm}{Cluster 0} & \includegraphics[height=.08\textheight]{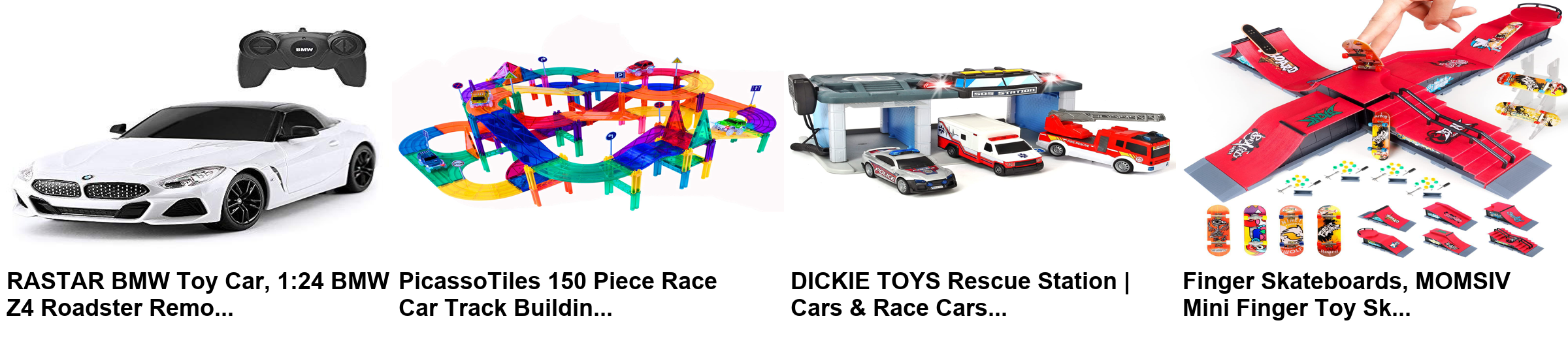} \\[0.1cm]
        \vspace{-1cm}{Cluster 2} & \includegraphics[height=.08\textheight]{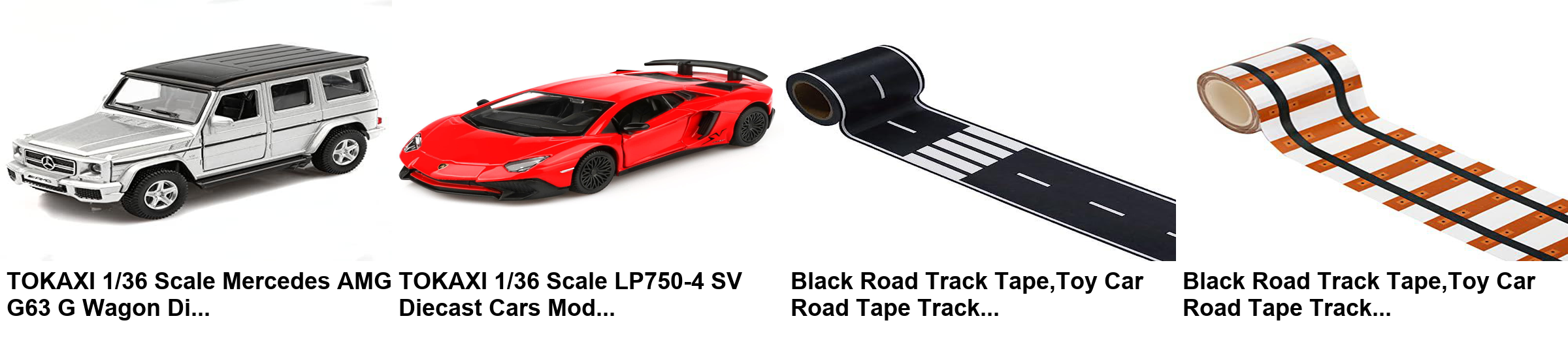} \\[0.1cm]
        \vspace{-1cm}{Cluster 4} & \includegraphics[height=.08\textheight]
        {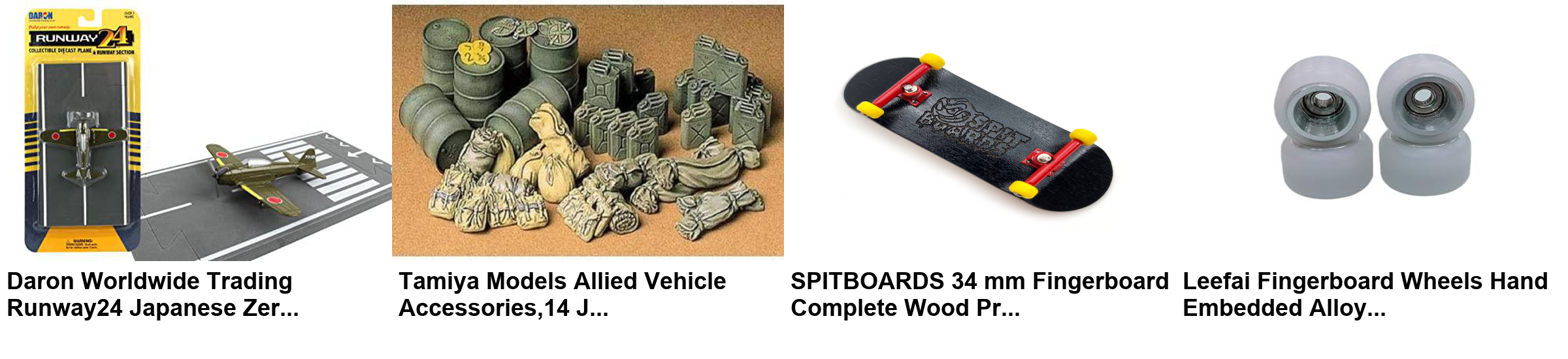} \\
    \end{tabular}
    \caption{Examples of closest products to cluster centers (Tabular + Text); Only three examples of clusters shown.}\label{fig:cluster_pca_txt_3d_combined_table_noimage}
\end{table}

\begin{table}[ht]\small
    \centering
    \footnotesize
    \begin{tabular}{@{}>{\centering\arraybackslash}m{0.15\textwidth} m{0.6\textwidth} m{0.2\textwidth}@{}}
       Cluster 
0 & Movie-Themed Character Collectibles and Customization Kits & \includegraphics[height=0.07\textheight]{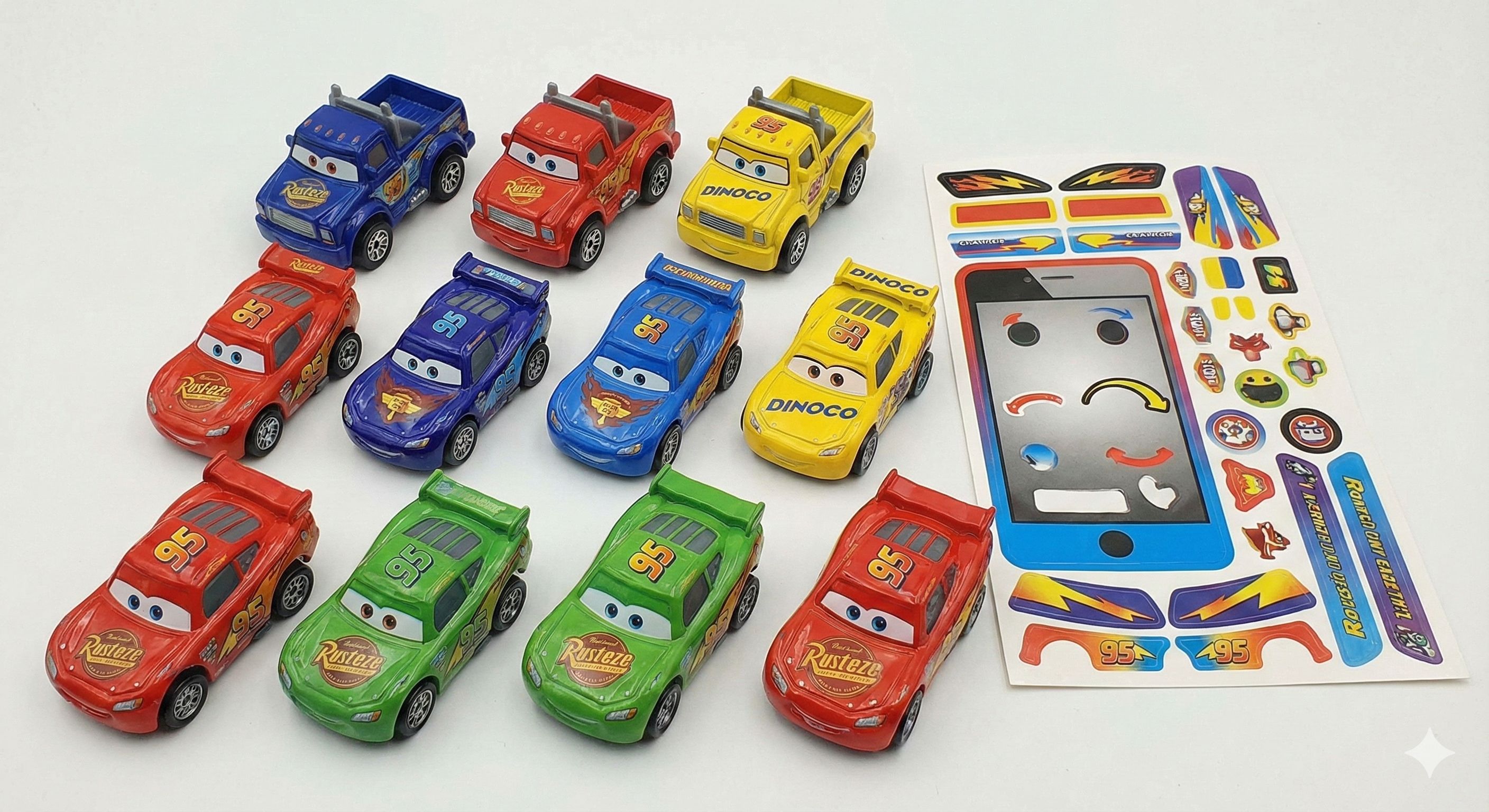} \\
        Cluster 1 & Realistic Large-Scale Utility Trucks with Functional Components & \includegraphics[height=0.07\textheight]{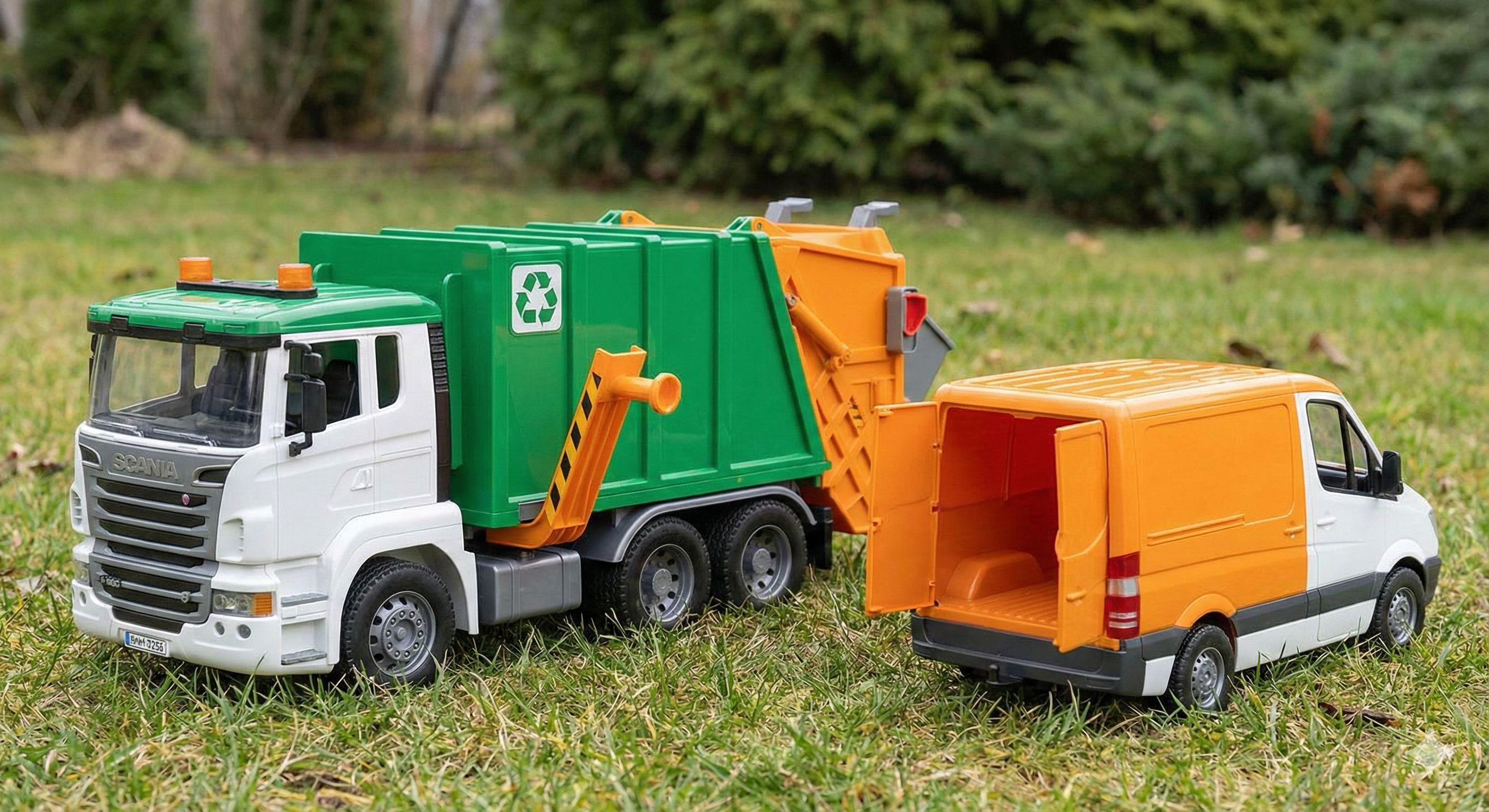} \\
        Cluster 2 & 1:64 Scale Die-Cast Real-World Automotive Replicas & \includegraphics[height=0.07\textheight]{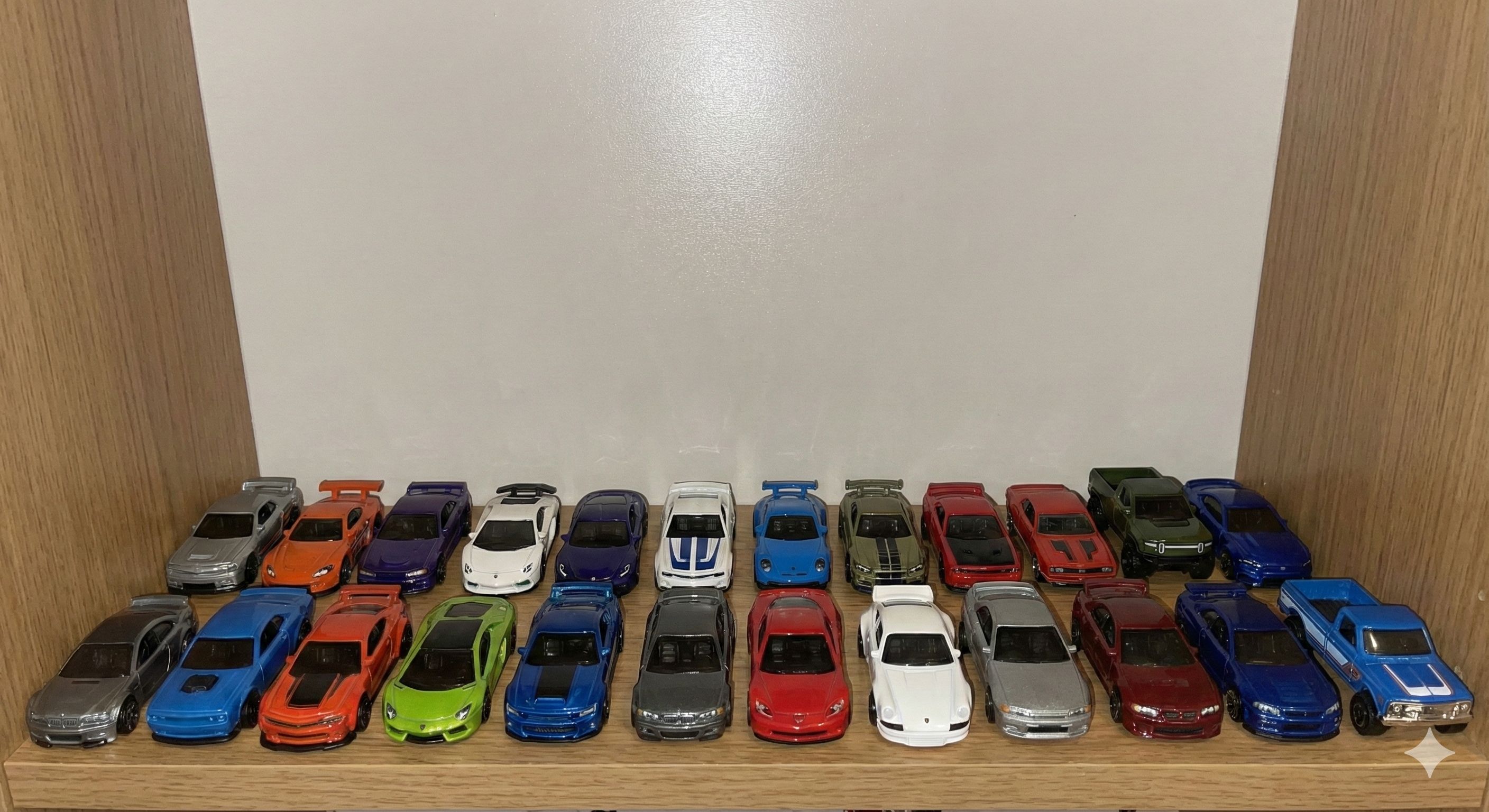} \\
        Cluster 3 & Medium-Scale Pull-Back Models and Wooden Train Sets & \includegraphics[height=0.07\textheight]{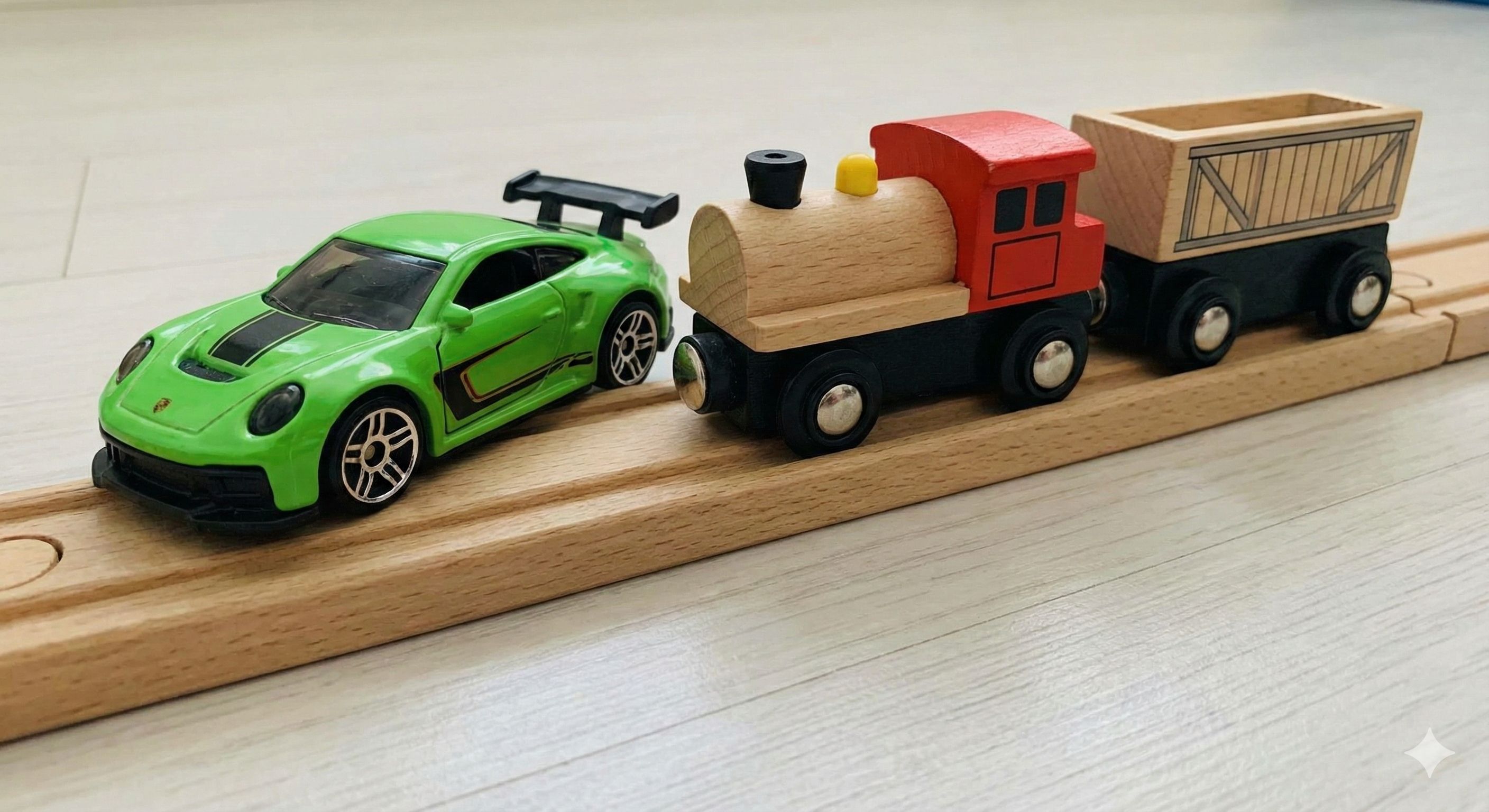} \\
        Cluster 4 & Interactive Stunt Vehicles, Launchers, and Adventure Playsets & \includegraphics[height=0.07\textheight]{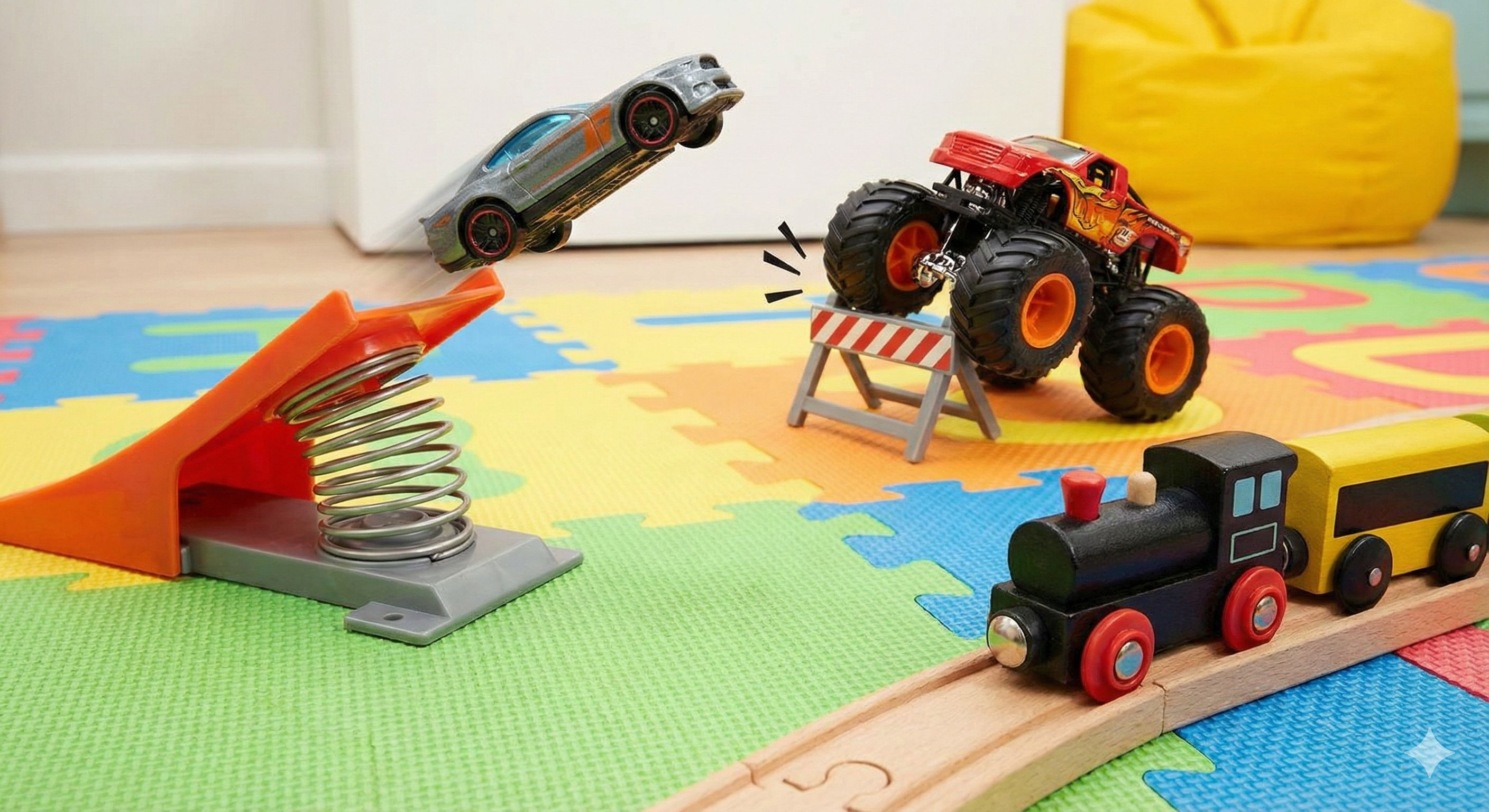} \\ 
    \end{tabular}
    \normalsize
    \caption{Generative AI summaries and images for the five cluster centroids (Tabular + Text + Image). }
\label{tab:gen_ai_summary}
\end{table}

\subsubsection{Quantitative Assessment}

While the previous discussion provides a qualitative indication that the model can represent products effectively, we now present a more quantitative assessment of the model’s predictive performance.

We begin by examining how well the embeddings---and their ``compressed'' versions---predict price and quantity levels, as well as their changes. Formally, our targets are \(Y \in \{Q, P, \Delta Q, \Delta P\}\). We also use these predictive regressions to fine-tune the embeddings themselves.

As shown in Table~\ref{tab:r2}, simple linear regressions using only tabular data perform poorly. Boosted trees yield substantial gains in predictive accuracy, and neural networks with text embeddings perform even better. Including image embeddings offers further improvements, though the additional gains are modest. Nonetheless, these gains are meaningful.

We also assess each model’s ability to predict changes in quantities and prices, rather than their levels. As expected, predicting changes is notably more difficult, leading to a sharp decline in performance. Even the best models achieve only about 15\% \(R^2\) for changes in quantity and 1.5\% \(R^2\) for changes in price. Despite this drop, the results underscore that AI-generated features provide strong predictive power for quantity levels and some improvement---albeit smaller---for predicting changes.

Next, we investigate whether ``compressed'' embeddings preserve the information in the full-dimensional embeddings. Specifically, we consider:
\begin{itemize}
    \item \textbf{Principal Components (PCA):}
    $
      X^{pc}_{i,k} := \gamma_k^T X^e_{i},
      \ 
      X^{pc}_{i} := (X^{pc}_{i,k})_{k=1}^K,
    $
    where \(\gamma_k\) is the \(k\)-th eigenvector of the covariance matrix of \(X^e_{i}\), corresponding to the \(k\)-th largest eigenvalue.

    \item \textbf{Centroid Similarities (CS):}
    $
      X^{sim}_{i,k} := c_k^T X^e_{i},
      \ 
      X^{sim}_{i} := (X^{sim}_{i,k})_{k=1}^K,
    $
    where \(c_k\) is the centroid (mean) of the \(k\)-th cluster identified by \(k\)-means.
\end{itemize}

These vectors capture how similar the embedding vectors are to principal axes of variation---either principal components or $k$-means cluster centroids---using cosine similarity. 
As shown in Table~\ref{tab:r2compressed}, using only five principal components or five centroid similarities can retain nearly all the prediction-relevant information contained in the original embeddings. In particular, when using a boosted tree, these compressed embeddings nearly match the performance of a deep neural network that uses the entire text and image inputs. This finding can simplify downstream tasks. In this paper, we rely primarily on centroid similarities, as they appear more interpretable than principal components for our application.

Overall, these results suggest that AI-derived embeddings greatly improve predictions of price and quantity levels, although they are less effective for predicting price changes. This has crucial implications for causal (price sensitivity) analysis, discussed below. Notably, our findings indicate that product embeddings tend to act more as \textit{effect modifiers} (i.e., determinants of elasticity) rather than \textit{confounders} of the causal relationship.



\begin{table}[t]\small
\centering
\caption{Test $R^2$ scores for predicting quantity and price signals.}
\label{tab:r2}
\begin{tabular}{lcc|cc}
\hline\hline
{Method [features] \textbackslash Target }  & {$Q_{it}$} & {$P_{it}$} & {$\Delta Q_{it}$} & {$\Delta P_{it}$} \\ \hline
{Linear Reg  [all tabular]} & 20.84\% & 15.82\% & 7.08\% &	0.64\%\\
{Boosted Trees Reg [all tabular] } & 47.51\% & 18.54\% & 15.13\% & 0.00\%\\
{Deep Learning Reg [text only; invariant tabular]} & 50.31\% & 65.74\% & 9.48\% & 1.24\%\\
{Deep Learning Reg [image and text; invariant tabular]} & 50.44\% & 67.26\% & 10.89\% & 1.15\%\\
{Deep Learning Reg [text only; all tabular]} & 60.65\% & 64.89\% & 14.48\% & 1.01\% \\
{Deep Learning Reg [image and text; all tabular]} & 59.53\% & 66.69\% & 12.17\% & 1.65\% \\
\hline
\hline
\end{tabular}
\vspace{2mm}

\caption*{ \footnotesize Note: All models are trained on a training set and scores are evaluated on a test set. Predictions use lagged values of time-varying controls. Negative $R^2$ values are set to $0$.}
\end{table}



\begin{table}[t]\small
\centering
\caption{Test $R^2$ Scores for ML methods using DL-based PCAs and similarities together with tabular controls.}
\label{tab:r2compressed}
\begin{tabular}{lcc|cc}
\hline
\hline
{Method [+ DL Features] \textbackslash Target } & {$Q_{it}$} & {$P_{it}$} & {$\Delta Q_{it}$} & {$\Delta P_{it}$}\\ \hline
{Linear Reg [+5 PCAs]} & 46.47\% & 60.63\% & 7.06\% & 0.64\%\\
{Linear Reg  [+ 5 Similarities]} & 47.27\% & 57.02\% & 7.07\% & 0.64\%\\
{Linear Reg  [+256 Embeddings]}  & 51.24\% & 62.14\% & 5.47\% & 0.00\% \\
{Boosted Trees Reg [+5 PCAs]} & 53.22\% & 64.36\% & 13.79\% & 0.00\%\\
{Boosted Trees Reg [+5 Similarities]} & 52.58\% & 62.53\% & 14.63\% & 0.00\%\\
{Boosted Trees Reg [+256 Embeddings]}  & 55.28\% & 65.64\% & 10.11\% & 0.00\% \\[5pt] \hline \hline
\end{tabular}
\vspace{2mm}

\caption*{\footnotesize Note: All models are trained on a training set and scores are evaluated on a test set. Predictions use lagged values of time-varying controls. Negative $R^2$ values are set to $0$.}
\end{table}

%% file: JASA_Buybox_Version/Section3.tex
\section{Estimating Price Effects} \label{sec:sensitivity}
Understanding how price changes affect consumers' choices is a central challenge in empirical economics and marketing. One common way to measure this relationship is through the elasticity of demand with respect to price. Although a regression of sales on prices may appear a straightforward way to estimate elasticity, it can yield biased estimates if key confounding factors are not properly accounted for. In this section, we explore various approaches to uncover the true causal price sensitivity and discuss their respective strengths and limitations.

\subsection{Initial Approach and Challenges}

A natural starting point is to estimate the relationship between price and product performance using a predictive model. Consider a regression of the (log) inverse ranking of product \(i\) at time \(t\), denoted \(Q_{it}\), on the (log) price \(P_{it}\) and a set of controls \(X_{it} = (X^e_{i}, X^o_{it})\), where \(X^o_{it}\) represents other tabular controls:
\[
\mathrm{L}[Q_{it} \mid P_{it}, X_{it}] = \delta P_{it} + g_t(X_{it}),
\]
where \(g_t(\cdot)\) is a function describing how the control variables influence the outcome over time. We allow \(g_t\) to vary with \(t\). The notation \(\mathrm{L}[Y \mid P,X]\) denotes the projection of the random variable \(Y\) onto the space of partially linear prediction rules of the form \(a\,P + g(X)\).

In practice, directly estimating this model often suggests a very small price sensitivity (or “elasticity”), captured by the coefficient \(\delta \in [-0.3, \, -0.1]\).
From a causal perspective, this result is implausible: the notion that a price change exerts virtually no effect on ranking or sales is counterintuitive, as it implies that raising prices would only marginally reduce the quantity sold. In other words, this finding suggests that certain key confounders—such as latent quality or visibility—are not adequately controlled for, resulting in a strongly biased estimate.

\subsection{The Causal Dynamic Model and Regressions} \label{sec:causal-model-dynamic}

To address the issue above, we introduce a simple dynamic panel data model to guide our statistical analysis. We can view the outcomes and key variables as arising from the following structural equation model (SEM).
\begin{assumption}(Structural Equation Model)
\begin{enumerate}
\renewcommand{\labelenumi}{\alph{enumi})}
\renewcommand{\theenumi}{\alph{enumi}}
\item \label{ass:sem-a}
It holds
\begin{eqnarray}\label{eq:causal1}
Q_{it} & = & a_t\bigl(S_{it}, \epsilon_{it}\bigr)\,P_{it} \;+\; q_t\bigl(S_{it}, \epsilon_{it}\bigr),\\[6pt]
P_{it} & = & p_t\bigl(S_{it}, \epsilon^p_{it}\bigr),\label{eq:causal2}\\[6pt]
S_{it} & = & s_t\bigl(S_{i,t-1}, \epsilon^s_{it}\bigr); \quad S_{it} \equiv (Q_{i,t-1}, P_{i,t-1}, X_{it}),\label{eq:causal3}
\end{eqnarray}
where \(a_t\), \(q_t\), \(p_t\), and \(s_t\) are nonparametric structural functions, and \(\epsilon_{it}\), \(\epsilon^p_{it}\), and \(\epsilon^s_{it}\) are i.i.d. stochastic vectors that are mutually independent.
\item \label{ass:sem-b}
The variables $Q_{it}$, $P_{it}$, and $S_{it}$ are square-integrable.
\end{enumerate}
\end{assumption}

This specification defines an \emph{autoregressive} model in which the quantity signal \(Q_{it}\) depends on the price signal \(P_{it}\) and other state variables \(S_{it}\). The state variables include lagged quantity and price, \(Q_{i,t-1}\) and \(P_{i,t-1}\), time-invariant product characteristics \(X_i\) (captured through embeddings), and time-varying characteristics \(X_{it}^o\) (such as ratings and the number of reviews). Among these variables, the lagged quantity \(Q_{i,t-1}\) is arguably a key confounder, reflecting both product visibility and quality—a conclusion reinforced by our empirical findings below. In particular, including the lagged quantity substantially shifts the estimated price elasticity into a more plausible range.

Because the model follows a Markovian structure, each period updates the state variables, after which prices and quantities respond to the new state vector. Figure~\ref{fig:dynamicSEM} illustrates this SEM graphically.

\begin{figure}[h]
\centering
\begin{tikzpicture} 
\node[font=\normalsize] (P) at (2,-2) {$P_{it}$};
\node[font=\normalsize] (S) at (0,0) {$S_{it}$};
\node[font=\normalsize] (Q) at (2,0) {$Q_{it}$};
\node[font=\normalsize] (Plag) at (-2,-2) {$P_{i,t-1}$};
\node[font=\normalsize] (Slag) at (-4,0) {$S_{i,t-1}$};
\node[font=\normalsize] (Qlag) at (-2,0) {$Q_{i,t-1}$};
\node[font=\normalsize] (Qlaglag) at (-6,0) {$\cdots$};
\node[font=\normalsize] (Qlead) at (4,0) {$S_{i,t+1}$};
\node[font=\normalsize] (Qleadlead) at (6,0) {$\cdots$};
\draw[thick, ->] (P) -- (Q);
\draw[thick, ->] (S) -- (Q);
\draw[thick, ->] (S) -- (P);
\draw[thick, ->] (Plag) -- (Qlag);
\draw[thick, ->] (Slag) -- (Qlag);
\draw[thick, ->] (Slag) -- (Plag);
\draw[thick, ->] (Plag) -- (S);
\draw[thick, ->] (Qlag) -- (S);
\draw[thick, ->, bend left=45] (Slag) to (S);
\draw[thick, ->, bend left=45] (S) to (Qlead);
\draw[thick, ->] (Q) -- (Qlead);
\draw[thick, ->] (P) -- (Qlead);
\end{tikzpicture}
\caption{A directed acyclic graph for the dynamic model.}\label{fig:dynamicSEM}
\end{figure}
It is convenient to define the variable $
A_{it} := a_t(S_{it}, \epsilon_{it}),$
which we interpret as a random elasticity or price sensitivity. Under this notation, the SEM above induces the potential outcomes \citep{Rubin1975bayesian}
\[
Q_{it}(p) = A_{it}\,p + q_t(S_{it}, \epsilon_{it}),
\]
by setting \(P_{it} = p\) in the first equation; see \cite{pearl1995causal}. Hence, the price sensitivity \(A_{it}\) is the causal effect of increasing \(P_{it}\) by one unit: $
\partial_p Q_{it}(p) = A_{it}.$
We focus on either the Average Causal Effect (ACE):
\[
\alpha_t = \mathrm{E}\bigl[\partial_p Q_{it}(p)\bigr] 
= \mathrm{E}[A_{it}],
\]
or the Conditional ACE (CACE):
\[
\alpha_t(S_{it}) = \mathrm{E}\bigl[\partial_p Q_{it}(p) \mid S_{it}\bigr] 
= \mathrm{E}[A_{it} \mid S_{it}],
\]
which describes the average causal effect conditional on product characteristics and thus captures the predictable component of price sensitivity.

Identification of both the CACE and the ACE in this setting follows from computing the conditional expectation of \(Q_{it}\) while conditioning on \(P_{it}\) (the treatment) and \(S_{it}\) (the observed confounders). Indeed, conditioning on \(S_{it}\) blocks non-causal sources of association between the outcome and the treatment \citep{pearl1995causal}. Including further lags of state variables $S_{it}$ is not strictly necessary under the Markovian structure, but it can serve as a useful specification check. Such identification assumes the absence of potential unobserved confounders, such as time-varying demand shocks that may affect both the outcome and the treatment. We further discuss limitations of our identification strategy and possible alternative strategies in Section 3.6, below.

We now derive the key regression function of interest:
\begin{equation}\label{eq:keyreg}
\mathrm{E}\bigl[ Q_{it} \mid P_{it}, S_{it} \bigr] 
= \alpha_t(S_{it}) \, P_{it} + \gamma_t(S_{it}),
\end{equation}
where $
\gamma_t(S_{it}) \;:=\; \mathrm{E}\bigl[q_t\bigl(S_{it}, \varepsilon_{it}\bigr){\color{red}{|\, S_{it}}}\bigr],$
so the CACE function \(\alpha_t(S_{it})\) appears as the heterogeneous slope in \eqref{eq:keyreg}. The ACE parameter \(\alpha_t\) then follows by averaging the CACE function over \(S_{it}\).

\subsection{Empirical Models}
In the empirical analysis, we examine two forms of the CACE function:
\begin{eqnarray}\label{eq:alpha HET}
& \textbf{I. Homogeneous Effect:}\quad
 & \alpha_t(S_{it}) = \alpha_t; \\[6pt]
&\textbf{II. Heterogeneous Effect:}\quad
& \alpha_t(S_{it}) 
= a_{0t} + \sum_{k=1}^K \alpha_{kt}\,X^{sim}_{i,k} \;+\; b_{1t}\,P_{i,t-1} \;+\; b_{2t}\,Q_{i,t-1}.
\end{eqnarray}
The first specification is very simple and serves as our baseline. The second is more elaborate yet still structured, allowing the elasticity function 
\(\;s \mapsto \alpha_t(s)\)
to depend on product characteristics as well as past quantities and prices:
\begin{itemize}
\item The first component of \(\alpha_t(s)\) captures product characteristics in the product space, represented by similarity vectors describing the product’s position.
\item The second part lets the elasticity vary with how popular the products are (lagged quantity) and how expensive they are (lagged price).
\end{itemize}
We show empirically that both components matter. In presenting our results, we assume time homogeneity by setting 
\(\alpha_t(\cdot) = \alpha(\cdot)\). Empirically, this did not affect any findings; we adopt this simplification purely for clarity of presentation.

For the model of the “control” function \(\gamma_t(S_t)\), we consider three cases:
\begin{itemize}
\item[1.] \textbf{Linear in State} \(S_{it}\): \(\;\gamma_t(S_{it}) = g_t^T\,S_{it}.\)
\item[2.] \textbf{Interactive Linear in State} \(S_{it}\): \(\;\gamma_t(S_{it}) = d_t^T\,I(S_{it}),\) where
\(I(S_{it})\) includes \(S_{it}\) and interactions of \(P_{i,t-1}\) and \(Q_{i,t-1}\) with \(X^{sim}_{it}\).
\item[3.] \textbf{Nonlinear in State} \(S_{it}\): \(\;\gamma_t(S_{it})\) is approximated by boosted trees.
\end{itemize}

The final model is fully nonparametric. We also experiment with using similarity vectors \(X_i^{sim}\) in place of the full 256-dimensional embedding \(X_i^e\) as controls, and find that the similarity vectors perform comparably well.

In summary, we will consider six types of empirical models, formed by the Cartesian product \(\{\mathrm{I, II}\} \times \{1,2,3\}\). Within each of these six types, we also vary how the control variables are included. As a preview of results, we note that the heterogeneous-effects model (II) receives the strongest empirical support, with variants of types 1–3 yielding similar quantitative results on elasticity.

\subsection{Orthogonal Inference of Causal Effects}
\label{subsec:orthogonal-inference}

We identify and estimate the causal effects using the following projection equation:
\begin{equation}
\label{eq:res}
 Q^\perp_{it} = \delta_t(S_{it}) P^\perp_{it} + e_{it},  \quad e_{it} \perp  P^\perp_{it} \;\mid\; S_{it},
\end{equation}
where \(e_{it} \perp  P^\perp_{it} \;\mid\; S_{it}\) means \(\mathrm{E}[e_{it} P^\perp_{it} \mid S_{it}] = 0\). The pair \(\bigl(Q^\perp_{it}, P^\perp_{it}\bigr)\) consists of the residuals
\[
Q^\perp_{it} =  Q_{it} - \mathrm{E}[Q_{it} \mid S_{it}],  
\quad 
P^\perp_{it} =  P_{it} - \mathrm{E}[P_{it} \mid S_{it}].
\]
The coefficient function \(\delta_t(S_{it})\) is the conditional predictive effect (CAPE) of a shock in the exposure variable on a shock in the outcome:
\[
\delta_t(S_{it}) := \frac{\mathrm{E}\bigl[Q^\perp_{it}\,P^\perp_{it}\mid S_{it}\bigr]}{\mathrm{E}\bigl[P^{\perp 2}_{it} \mid S_{it}\bigr]}.
\]
Averaging the CAPE gives the average predictive effect (APE), \( \mathrm{E}[ \delta_t(S_{it}) ]\).

We now make the following observation.

\begin{prop}[Identification of CACE]
If Assumption~1(\ref{ass:sem-a}) and (\ref{ass:sem-b}) hold, then the CACE is identified by the CAPE:
\[
\alpha_t(S_{it}) = \delta_t(S_{it}),
\]
almost surely, provided that both exist and are finite. Then the ACE is identified by the APE, \(\mathrm{E}\bigl[\alpha_t(S_{it})\bigr] = \mathrm{E}\bigl[\delta_t(S_{it})\bigr]\), again provided these expectations exist and are finite.
\end{prop}


\begin{proof}
Note that from equation \eqref{eq:keyreg} we have $Q_{it} = \alpha_t(S_{it})P_{it} + \gamma_t(S_{it}) + u_{it}$, where $u_{it} := q_t(S_{it},\varepsilon_{it}) - \gamma_t(S_{it})$ satisfies $\mathrm{E}[u_{it} | S_{it}] = 0$ by construction. Subtracting $\mathrm{E}[-|S_{it}]$ from both sides gives $Q_{it}^{\perp} = \alpha_t(S_{it})P_{it}^{\perp} + u_{it}$. Assumption 1 ensures that $\mathbb{E}[u_{it} P_{it}^{\perp} | S_{it}] = 0$, since $u_{it}$ depends only on $(S_{it}, \varepsilon_{it})$ while $P_{it}^{\perp}$ depends only on $(S_{it}, \varepsilon_{it}^p)$, and the error terms are mutually independent. Multiplying both sides by $P_{it}^\perp$ and taking conditional expectations w.r.t.~$S_{it}$ then gives
\(\mathrm{E}[Q_{it}^\perp P_{it}^\perp|S_it] = \alpha_{it}(S_{it})\mathrm{E}[(P_{it}^\perp)^2|S_{it}] \), hence $\delta_{t}(S_{it}) = \alpha_t(S_{it})$.
\end{proof}
It follows that once we learn the CAPE, we effectively learn the CACE, provided that the causal SEM postulated above holds. If the SEM is only approximate, then the CAPE can still be treated as an approximation to the CACE; we elaborate on this in Section~\ref{sec:crit}.

We estimate the CAPE and CACE using both linear regression and nonlinear, nonparametric models, applying modern machine-learning tools and cross-fitting to compute the residualized outcomes and exposure variables. We then estimate the projection equation~\eqref{eq:res} using either homogeneous or heterogeneous forms of \(\delta_t(\cdot) = \alpha_t(\cdot)\) via least squares, and apply conventional statistical inference to construct \(p\)-values and confidence intervals, following \cite{chernozhukov2018}. Details on the workflow and the algorithms used are provided in the Online Appendix.

\begin{remark}[Orthogonalization]
The argument above relies on the classical partialling-out or orthogonalization approach \citep{frisch1933partial,lovell1963seasonal,robinson1988root}. 
The “residual-on-residual” method underlies double machine learning (also called $R$-learning), which uses cross-fitted machine learning to estimate residuals and then infers the CAPE by least squares \citep{chernozhukov2018,nie2021quasi,semenovaR}. 
This approach is part of a broader class of debiased machine learning (DML) algorithms rooted in semiparametric learning theory \citep{levit1975efficiency,hasminskii1978,pfanzagl:wefelmeyer}.\qed
\end{remark}

\begin{remark}[Causal Fine-Tuning]
Recall that we fine-tuned the embeddings \(X_i^{e}\) to produce the best-performing prediction rules for \(Q_{it}\) and \(P_{it}\) (or their temporal differences, since past values strongly predict future outcomes). In this sense, the fine-tuning was well-suited to our causal inference problem. This idea generalizes to fully nonlinear models, where one could fine-tune embeddings to learn the Neyman-orthogonal equations for the parameter of interest.\qed
\end{remark}

\begin{remark}[Robustness of DML to Estimation Noise in Embeddings]
One might suspect that using estimated embeddings rather than “optimal” ones would complicate inference. However, under mild conditions, this is not the case. The key defining property of DML is that its estimating equations are robust to perturbations in the nuisance regression function—a feature referred to as Neyman orthogonality. Since perturbations in regressors translate to perturbations in the regression function, the former produce a \textit{zero} first-order effect on the DML estimator. We provide further theoretical details on this point in Addendum~A.\qed
\end{remark}

\subsection{Empirical Results}

\subsubsection{Homogeneous Elasticity Model} 
We begin by examining the homogeneous elasticity function. Table~\ref{tab:dynamic_effect_analysis} shows that, in various specifications—ranging from simpler linear regressions with lagged price and quantity to more complex boosted-tree models incorporating product embeddings and additional controls—price consistently exerts a negative and highly significant effect on the quantity signal (the inverse sales rank). The confidence intervals range from about \(-0.79\) to \(-0.54\), indicating strong economic and statistical significance. Indeed, these estimates imply that a 1\% price increase reduces the inverse sales rank by about \([0.54, 0.79]\%\). Furthermore, to convert this effect into an elasticity of the actual demand under the power law assumption (cf. Remark~\ref{rmk:time-avg}), we need to multiply these coefficients by about 2, which indicates that the demand elasticity itself falls in the range of approximately \([-1.58, -1.08]\).
\begin{table}[t]
\centering
\caption{Estimated price effects based on the partially linear dynamic model.}
\label{tab:dynamic_effect_analysis}
\resizebox{1\textwidth}{!}{
\begin{tabular}{lcccccc}
\hline\hline
  Specification of Control Function (State $S_{t}$)             & {coef} & {std err} & {t} & {P-val.} & {[5.0\%, } & {95.0\%]} \\ \hline
{I-1. Linear ($P_{t-1}$, $Q_{t-1}$)}                        & -0.690 & 0.040 & -17.248 & \textless 0.001 & -0.756 & -0.624\\ 
{I-1. Linear ($P_{t-1}$, $Q_{t-1}$, $X^{e}, X^{o}_t$)}      & -0.712 & 0.039 & -18.364 & \textless 0.001 & -0.776 & -0.649\\ 
{I-1. Linear ($P_{t-1}$, $Q_{t-1}$, $X^{sim}, X^{o}_t$)}    & -0.723 & 0.039 & -18.725 & \textless 0.001 & -0.786 & -0.659 \\
{I-2. Linear with Interactions ($P_{t-1}$, $Q_{t-1}$, $X^{sim}, X^{o}_t$)}    & -0.727 & 0.039 & -18.698 & \textless 0.001  & -0.791 & -0.662\\
{I-3. Boosted Trees ($P_{t-1}$, $Q_{t-1}$, $X^{e}, X^{o}_t$)}    &  -0.697 & 0.049 & -14.362 & \textless 0.001 & -0.704 & -0.542\\
{I-3. Boosted Trees ($P_{t-1}$, $Q_{t-1}$, $X^{sim}, X^{o}_t$)}    & -0.691 & 0.041 & -17.051 & \textless 0.001 & -0.777 & -0.617\\
\hline\hline
\end{tabular}}
  \caption*{\footnotesize Note: Standard errors are clustered at the product level. The LR models are estimated using OLS. The PLR model is estimated using DML with cross-fitted boosted trees. }
\end{table}

The results consistently suggest that incorporating lagged quantities and prices—which capture a product’s popularity and quality—reveals a more substantial, economically meaningful negative price elasticity. Interestingly, other product characteristics, including embeddings and similarities, are not statistically or economically significant confounders here. This finding is consistent with the results in Section~\ref{sec-repr-products}, which indicate that product embeddings or similarities do not strongly predict price changes. Hence, including or excluding these controls does not substantially alter the estimates. However, this does not rule out the possibility that product embeddings could be crucial \emph{effect modifiers}, as we discuss next.

\subsubsection{Heterogeneous Elasticity Model}

\begin{figure}[ht]
    \centering
    \includegraphics[width=.7\linewidth,height=.45\linewidth]{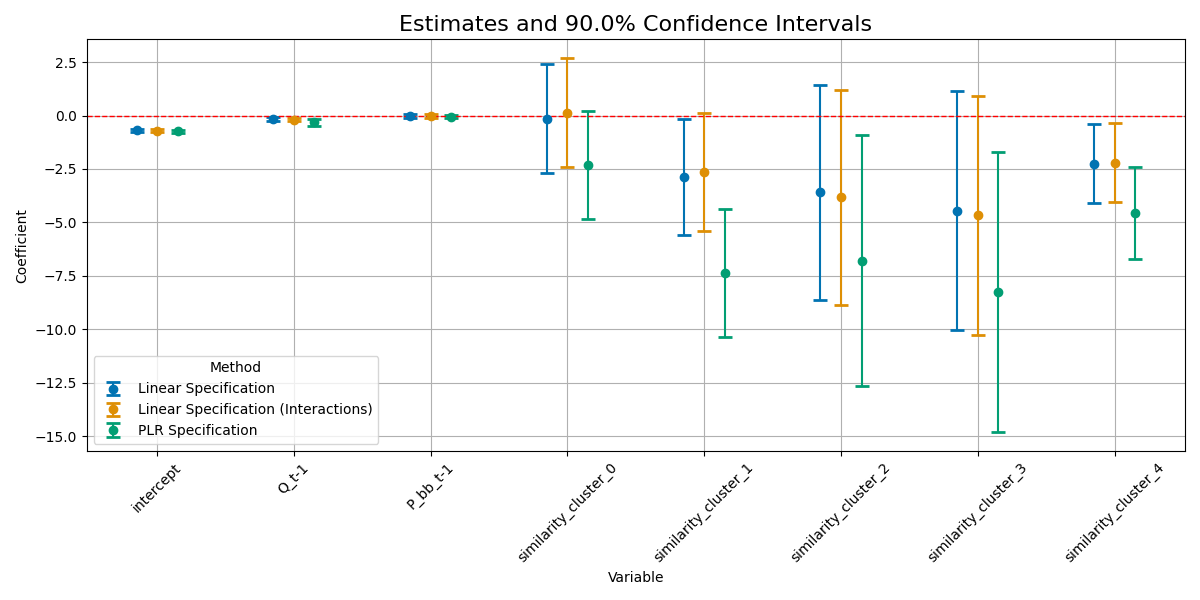}
    \caption{Parameters of the Estimated Elasticity Function and 90\% Confidence Intervals}
    \caption*{\footnotesize Note: Covariance is clustered at the product level. The LR model is estimated using OLS. The PLR model is estimated using DML with cross-fitted boosted trees. }
    \label{fig:cluster_similarity_coefs_dynamic}
\end{figure}

While the homogeneous specification yields more plausible estimates of average price sensitivity, it may still obscure important differences across products. The next step, therefore, models heterogeneous effects, allowing price elasticity to vary as a function of observed product characteristics or embedding-based similarity measures. Figure~\ref{fig:cluster_similarity_coefs_dynamic} and Table \ref{tab:dynamic_effect_analysis_cate_PLR} illustrate how the elasticity estimates relate to these cluster similarities and to lagged price and quantity signals. These results reveal that certain products are more price-sensitive than others. For example, higher-priced and better-ranked products show greater price sensitivity.\footnote{This finding is most clearly seen in Table \ref{tab:dynamic_effect_analysis_cate_PLR}; note that the negative coefficient on lagged price lacks statistical significance, while the negative coefficient on lagged sales rank is highly significant.} Similarities to cluster centroids also emerge as important drivers of sensitivity. Products associated with specific clusters experience notably different price-sensitivity levels, underscoring the importance of accounting for product-level heterogeneity when estimating price effects.

\begin{table}[t]
\small
\centering
\caption{Inference on the price effect modifiers with nonlinear (Boosted Trees) state control (Model II-3). Results for Models II-1 and II-2 are similar.}
\label{tab:dynamic_effect_analysis_cate_PLR}
\resizebox{.7\textwidth}{!}{
\begin{tabular}{lcccccc}
\hline\hline
   Modifier              & {coef} & {std err} & {t} & {P-val.} & {[5.0\%, } & {95.0\%]} \\ \hline
{Centercept} & -0.736 & 0.046 & -15.860 & \textless 0.001 & -0.812 & -0.659 \\
{Lagged Quantity, $Q_{t-1}$} & -0.325 & 0.092 & -3.515 & \textless 0.001 & -0.477 & -0.173 \\
{Lagged Price, $P_{t-1}$} & -0.049 & 0.035 & -1.401 & 0.161 & -0.106 & 0.008 \\
{Cluster Similarity $0$} & -2.312 & 1.524 & -1.517 & 0.129 & -4.819 & 0.195 \\
{Cluster Similarity $1$} & -7.365 & 1.831 & -4.023 & \textless 0.001 & -10.377 & -4.354 \\
{Cluster Similarity $2$} & -6.785 & 3.558 & -1.907 & 0.056 & -12.637 & -0.933 \\
{Cluster Similarity $3$} & -8.255 & 3.985 & -2.072 & 0.038 & -14.810 & -1.700 \\
{Cluster Similarity $4$} & -4.559 & 1.303 & -3.500 & \textless 0.001 & -6.702 & -2.417 \\
\hline\hline
\end{tabular}}
\caption*{\footnotesize Note: Standard errors are clustered at the product level. The LR model is estimated using OLS. The PLR model is estimated using DML with cross-fitted boosted trees. Lagged quantities and prices are centered and rescaled to have unit variance across \((i,t)\).  The coefficient on the centercept represents the average effect. }
\end{table}

\begin{figure}[ht]
    \centering
    \includegraphics[width=1\linewidth,height=.5\linewidth]{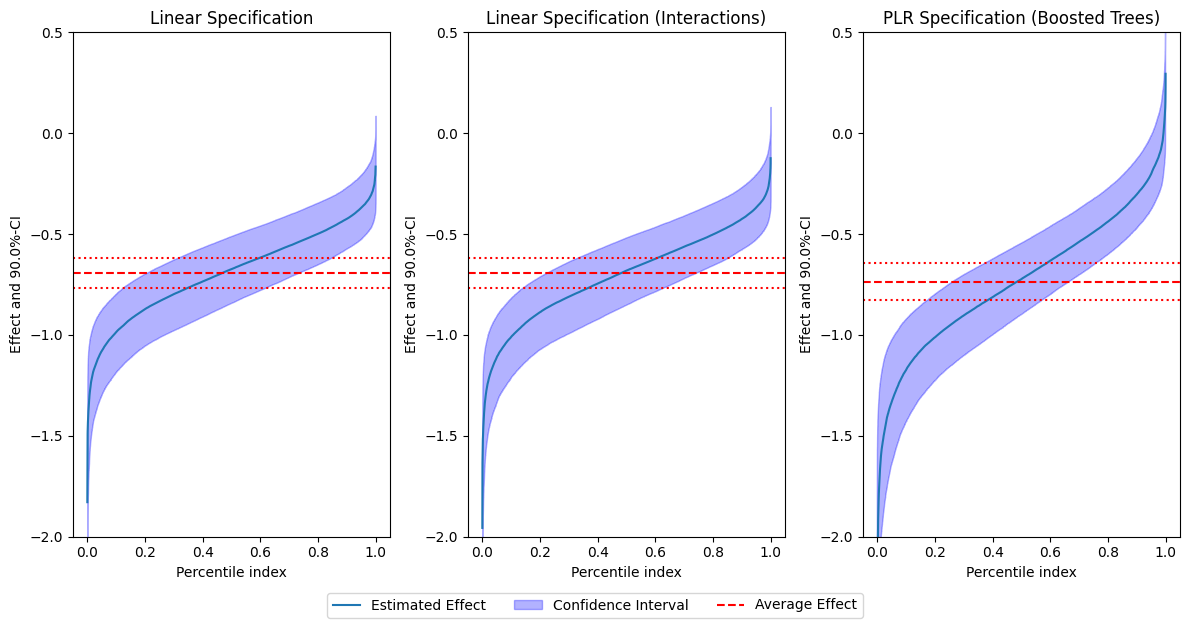}
    \caption{Sorted Elasticity as a Function of Effect Modifiers and 90\% Pointwise Confidence Bands }
    \caption*{\footnotesize Note: Covariance is clustered at the product level. The LR model is estimated using OLS. The PLR model is estimated using DML with cross-fitted boosted trees. }
    \label{fig:cluster_similarity_sorted_effects_dynamic}
\end{figure}

To provide a comprehensive view of how strong this heterogeneity can be, Figure~\ref{fig:cluster_similarity_sorted_effects_dynamic} plots each product’s estimated elasticity (vertical axis) sorted in ascending order (horizontal axis is the “percentile index”); as per \cite{chernozhukov2018sorted}. The solid blue line represents the point estimates, the blue band denotes the 90\%  pointwise confidence intervals, and the dashed red line indicates the overall average elasticity. Reflecting the discussion of heterogeneous price sensitivity, elasticity ranges from about \(-2.0\) to nearly \(0.4\), and these wide differences are statistically significant (as shown by the confidence bands). By multiplying these numbers by 2, we estimate that actual demand elasticity spans roughly \(-4.0\) to \(0.8\), with an average near \(-1.5\). This range highlights the importance of allowing for heterogeneity in price responses, as some products appear far more (or far less) sensitive to price changes than the average.

Tests of joint significance further confirm that cluster similarities matter. Table~\ref{tab:cluster_significance_dynamic} presents \(p\)-values from a \(\chi^2\)-test that evaluates whether the cluster similarity measures jointly have a statistically significant impact on price sensitivity. The results indicate that these cluster-based heterogeneities are  significant, reinforcing the conclusion that certain products are more responsive to price changes than others.

\begin{table}[ht]\footnotesize
    \centering
    \caption{$p$-values for $\chi^2$-test of joint significance of price effect modifiers}
    \label{tab:cluster_significance_dynamic}
    \begin{tabular}{@{}lcc@{}}
        \hline\hline
        Model & All Modifiers  &   Similarities Only\\ \midrule
        II.1 Linear Specification & \textless 0.001 & 0.030 \\
        II.2 Linear Specification (Interactions) & \textless 0.001 & 0.054 \\
        II.3 PLR Specification (Boosted Trees) & \textless 0.001 & 0.001 \\ \hline\hline
    \end{tabular}
    \caption*{\footnotesize Note: Standard errors are clustered at the product level. LR is estimated using OLS. The PLR model is estimated using DML with cross-fitted boosted trees (sample size: $38,041$).}
\end{table}

\subsection{Limitations of the Analysis and Sensitivity Bounds}\label{sec:crit}

Our statistical estimates carry a well-defined causal interpretation under the stated, relatively strong assumptions. The main threat to this interpretation is the presence of latent, time-varying factors that bias the relationship between \(P_{it}\) (the price signal) and \(Q_{it}\) (the quantity signal). Indeed, we might suspect that price endogenously responds to demand shocks \(\epsilon_{it}\) from the outcome equation, effectively making \(\epsilon_{it}\) a confounder.
The DAG below illustrates such a scenario. 

However, in our setting, we suspect that the link \(\epsilon_{it} \to P_{it}\) may be weak, as prices often follow “sticky,” piecewise-constant paths that do not change as frequently as quantity signals (sales ranks); see, for example, Figure~\ref{fig:example} in Section \ref{sec-repr-products}. Formally, if the edge from \(\epsilon_{it}\) to \(P_{it}\) is zero, our earlier identification strategy holds, and our estimates are indeed causal. Otherwise, we can treat them as approximations of the causal estimates. 

For methods to bound the effect distortion caused when the omitted confounder \(\epsilon_{it}\) affects \(P_{it}\), we can employ the the senstitivity framework of \cite{chernozhukov2021long}. Section 4 of the Online Appendix uses this framework to investigate how strong unobserved confounding would need to be to materially alter the estimated price effect. For concreteness, we focus on the nonlinear (boosted-trees) state-control estimate of the rank elasticity, -0.697 (Model I-3), and consider the adversarial case in which the omitted factor shifts demand and price in perfectly aligned directions so as to maximize bias. Allowing a latent confounder to explain up to 5\% of the residual variation in both quantity and price yields an estimated rank-elasticity bound of [-0.919, -0.475], with corresponding $90\%$ one-side confidence bounds of -0.979 and -0.409, placing the effect squarely within the economically plausible range for rank elasticities. To obtain bounds wide enough to include zero while also accommodating very large elasticities in magnitude (near -1.5), the omitted confounder would need to explain roughly 15\% of the residual variation in both outcome and price. This requirement is demanding relative to observed benchmarks: after adjusting for lagged outcomes and lagged prices, which proxy the relevant demand states, the remaining observed covariates contribute essentially no explanatory power for residual movements in either price or quantity, making the 5\% adversarial-confounding benchmark already conservative.

\begin{figure}[H]
\centering
\begin{tikzpicture} 
\node[font=\normalsize] (P) at (2,-2) {$P_{it}$};
\node[font=\normalsize] (S) at (0,0) {$S_{it}$};
\node[font=\normalsize][draw, circle, dashed] (eps) at (0,-2) {$\epsilon_{it}$};
\node[font=\normalsize] (Q) at (2,0) {$Q_{it}$};
\node[font=\normalsize] (Plag) at (-2,-2) {$P_{i,t-1}$};
\node[font=\normalsize] (Slag) at (-4,0) {$S_{i,t-1}$};
\node[font=\normalsize] (Qlag) at (-2,0) {$Q_{i,t-1}$};
\node[font=\normalsize][draw, circle, dashed] (epslag) at (-4,-2) {$\epsilon_{i,t-1}$};
\node[font=\normalsize] (Qlaglag) at (-6,0) {$\cdots$};
\node[font=\normalsize] (Qlead) at (4,0) {$S_{i,t+1}$};
\node[font=\normalsize] (Qleadlead) at (6,0) {$\cdots$};
\draw[thick, ->] (eps) -- (Q);
\draw[thick, ->] (eps) -- (P);
\draw[thick, ->] (P) -- (Q);
\draw[thick, ->] (S) -- (Q);
\draw[thick, ->] (S) -- (P);
\draw[thick, ->] (Plag) -- (Qlag);
\draw[thick, ->] (epslag) -- (Qlag);
\draw[thick, ->] (epslag) -- (Plag);
\draw[thick, ->] (Slag) -- (Qlag);
\draw[thick, ->] (Slag) -- (Plag);
\draw[thick, ->] (Plag) -- (S);
\draw[thick, ->] (Qlag) -- (S);
\draw[thick, ->, bend left=45] (Slag) to (S);
\draw[thick, ->, bend left=45] (S) to (Qlead);
\draw[thick, ->] (Q) -- (Qlead);
\draw[thick, ->] (P) -- (Qlead);
\end{tikzpicture}
\caption{Dynamic model with demand shocks as omitted confounders.}\label{fig:dynamicSEMshocks}
\end{figure}

Another way to achieve identification is through an instrumental variable \(Z_{it}\) that induces exogenous variation in \(P_{it}\), independent of \(\epsilon_{it}\). By conditioning on the variation produced by \(Z_{it}\), we isolate a component of \(P_{it}\) uncorrelated with \(\epsilon_{it}\), which then identifies the causal effect of price changes on quantity changes. Essentially, one can estimate the average causal effect of \(Z_{it}\) on \(Q_{it}\) (controlling for \(S_{it}\)), and the average causal effect of \(Z_{it}\) on \(\Delta P_{it}\) (also controlling for \(S_{it}\)), and then take their ratio—following the classical approach of Philip Wright (1928), republished as \cite{wright:1928}—to identify the causal effect of \(P_{it}\) on \(Q_{it}\). We could not find such credible instruments in our setting. Using further lagged values of \(Q_{i,t-1}\) is one potential approach, but it would require taking the first-order autoregressive specification  too literally (i.e., those further lagged variables should likely be included as state variables, rather than employed as instruments).

%% file: JASA_Buybox_Version/Conclusion.tex
\section{Concluding Remarks} \label{sec:conclusion}
This study highlights the significant potential of AI-generated multimodal embeddings in demand analysis. By integrating text, image, and tabular data into a causal econometric framework, we improve both the precision of demand and price forecasts and the credibility of elasticity estimates. Our findings show that these rich embeddings not only enhance predictive accuracy but also reveal substantial heterogeneity in consumer price sensitivity, providing nuanced insights into demand behavior. These advancements create a methodological bridge between machine learning and econometrics, illustrating how modern AI tools can enrich traditional economic analyses. 

\section*{Acknowledgments and Data Availability} We would like to thank the anonymous Associate Editor and a referee, and also Bulat Gafarov, Emi Nakamura, Patrick Kline, and Sylvia Klosin,  for useful comments that improved the paper.  We also thank the AI tools -- Refine.Ink and ChatGPT-Pro -- for additional helpful comments that we incorporated during the revision. Data and code are available at \\ \texttt{https://github.com/JanTeichertKluge/demand-analysis-repro}.

%% file: JASA_Buybox_Version/Section4.tex
\section{Addendum:  Robustness to Estimated Embeddings} 
One may suspect that using estimated embeddings instead of "optimal" ones could complicate inference. However, under mild conditions, this is not the case. To illustrate this point in a simple manner, consider the homogeneous model:
$$
Q_{it}^\perp = \delta_t P_{it}^\perp + e_{it}, \quad e_{it} \perp P_{it} \mid S_{it}.
$$

Let $Y_{it}$ denote the predictive target, which is either $P_{it}$ or $Q_{it}$. Let $X^e_i(\hat{\phi})$ denote the estimated and fine-tuned embeddings, where $\hat \phi$ denotes estimated parameters, and \(S_{it}(\hat{\phi})\) the derived controls. We assume that $\hat \phi$ is obtained from data that are independent of the main data used in the analysis. Similarly, let $X_i^e$ and $S_{it}$ denote the \emph{ideal} embeddings and controls, in the sense that
\[
\mathrm{E}[Y_{it} \mid  S_{it}, S_{it}(\hat \phi) ] = \mathrm{E}[Y_{it} \mid  S_{it}].
\]
In other words, after including \(S_{it}\), the best prediction rule for \(Y_{it}\) given both \(S_{it}\) and \(S_{it}(\hat \phi)\) depends only on $S_{it}$.

Consider a learner $\hat \gamma^Y_{t} (S_{it}(\hat{\phi}))$ that
minimizes empirical risk
\(
\sum_{i \in A} \{Y_{it} - \gamma(S_{it}(\hat{\phi}))\}^2 
\)
over control functions $\gamma$ in the convex model $\mathcal F$, conditional upon fine-tuned embeddings $\hat{\phi}$ (this includes our considered models 1-3). Here $A$ is a subset of $\{1,..,n\}$ whose size is proportional to $n$. The DML inference approaches using ideal and estimated embeddings are first-order equivalent under the following two key conditions:
\begin{itemize} 
\item[(E1)] For the given $\hat\phi$, the \textit{square root} of the offset Rademacher complexity \citep{liangrakhlin:ORC} of the class $\mathcal{F}_{\hat{\phi}} = \{\gamma(S_{it}(\hat{\phi})): \gamma \in \mathcal F\}$ is $o_p(n^{-1/4})$.

\item[(E2)] The approximation error of the model $\mathcal{F}$ with estimated embeddings $\hat \phi$ is sufficiently small:  
$
\inf_{\gamma \in \mathcal{F}} \sqrt{\mathrm{E}[ \{ \mathrm{E}[Y_{it} \mid S_{it}] -  \gamma( S_{it}(\hat{\phi})) \}^2 \mid \hat{\phi} ]} = o_p(n^{-1/4}).
$
\end{itemize}

For condition (E1), it is useful to recall that for high-dimensional parametric models with $d$ parameters, the square root of the offset complexity scales as $\sqrt{d/n}$, so the condition above requires the dimension $d$ is $o(\sqrt{n})$; see \citet{liangrakhlin:ORC} and \citet{bach:book} for further discussion and bounds for other classes of nonparametric learners.
Condition (E2) depends on the specification of the model $\mathcal{F}$ as well as the quality of the fine-tuned embeddings $\hat \phi$. It also tells us that fine-tuning should be done with the goal of predicting the labels $Y_{it}$; the better we do this, the more plausible condition (E2) becomes.

\begin{prop}Work with the setup in this section. Assume that the data $W_{it} = (Q_{it}, P_{it}, X^{in}_{it})$, 
are identically distributed across $i$ for  each $t =1,..,T$, where $T$ is fixed. 
 Assume conditions (E1) and (E2) hold and that $Y_{it}$ and $\mathcal{F}$ are bounded. Then the empirical risk minimizer $\hat \gamma^Y_t (S_{it} (\hat \phi))$ learns 
$\mathrm{E}[Y_{it} \mid S_{it}]$ at the rate  \( o_p(n^{-1/4}) \):
$\max_t \sqrt{\mathrm{E}[ \{ \mathrm{E}[Y_{it} \mid S_{it}] -  \hat \gamma( S_{it}(\hat{\phi})) \}^2 \mid \hat{\phi} ]} = o_p(n^{-1/4}).
$
\end{prop}

\begin{proof}
First note that for any $\gamma \in \mathcal{F}$, the random variable $\hat{Z}_{it}(\gamma) \coloneqq \gamma(S_{it}(\hat{\phi}))$ belongs to the subspace  $E \subset L^2$ consisting of functions that are measurable w.r.t.~the pair \( (S_{it}, S_{it}(\hat{\phi})) \). Thus we have 
\begin{equation}\label{eq:target-ideal} \mathrm{E}\{Y_{it} - \hat{Z}_{it}(\gamma)\}^2 = \mathrm{E}\{Y_{it} - \mathrm{E}[Y_{it}|S_{it}, S_{it}(\hat{\phi})]\}^2 + \mathrm{E}\{\mathrm{E}[Y_{it}|S_{it}, S_{it}(\hat{\phi})] - \hat{Z}_{it}(\gamma)\}^2
\end{equation}
by the Pythagorean theorem, since $\mathrm{E}[Y_{it}|S_{it}, S_{it}(\hat{\phi})]$ is the orthogonal projection of $Y_{it}$ onto $E$. It follows that the left-hand side depends on $\gamma$ only through:
\(\mathrm{E}\{\mathrm{E}[Y_{it}|S_{it}, S_{it}(\hat{\phi})] - \hat{Z}_{it}(\gamma)\}^2 = \mathrm{E}\{\mathrm{E}[Y_{it}|S_{it}] - \hat{Z}_{it}(\gamma)\}^2\), where we use the fact that $S_{it}$ is derived from the ideal embeddings.

Now, let \(\gamma^*_t\) in the closure of \(\mathcal{F}\) satisfy
\(\mathrm{E}\{Y_{it} - \hat{Z}_{it}(\gamma^*_t)\}^2 = \inf_{\gamma \in \mathcal{F}} \mathrm{E}\{Y_{it} - \hat{Z}_{it}(\gamma)\}^2\). By \eqref{eq:target-ideal} above, the same $\gamma^*_t$ also satisfies \(\mathrm{E}\{\mathrm{E}[Y_{it}|S_{it}] - \hat{Z}_{it}(\gamma^*_t)\}^2 = \inf_{\gamma \in \mathcal{F}} \mathrm{E}\{\mathrm{E}[Y_{it}|S_{it}] - \hat{Z}_{it}(\gamma)\}^2\), which is \(o_p(n^{-1/2})\) by (E2). 

Next, by Theorem 3 of \citet{liangrakhlin:ORC} combined with Markov's inequality, 
\(
\mathrm{E}\{Y_{it} -  \hat{Z}_{it}(\hat \gamma^Y_{t})\}^2
 \le \mathrm{E}\{Y_{it} - \hat{Z}_{it}(\gamma^*_t)\}^2 + o_p(n^{-1/2}).
 \)
    Decomposing both expectations using \eqref{eq:target-ideal}, we recover 
 \[
\mathrm{E}\{\mathrm{E}[Y_{it}|S_{it}] -  \hat{Z}_{it}(\hat \gamma^Y_{t})\}^2
 \le \mathrm{E}\{\mathrm{E}[Y_{it}|S_{it}] - \hat{Z}_{it}(\gamma^*_t)\}^2 + o_p(n^{-1/2}) = o_p(n^{-1/2})
 \]
 where the last step follows by the previous paragraph.
To conclude, we take a union bound over finitely many periods $t$ to deduce that 
 \(
\max_{t} \mathrm{E}\{\mathrm{E}[Y_{it}|S_{it}] -  \hat{Z}_{it}(\hat \gamma^Y_{t})\}^2
= o_p(n^{-1/2}),
\)
as needed.
\end{proof}

We describe the DML slope estimator next. Let $(I_\ell)_{\ell}^L$ be the partition of $[n]=\{1,..., n\}$ into $L$ folds of approximately equal size.  In step 1, for each $\ell$: obtain 
$\hat \gamma^Y_{t, \ell} (S_{it}(\hat \phi))$, the empirical risk minimizer over observation indices  $A =[n]\setminus I_\ell$, for $Y= Q$ and $Y =P$; then obtain the residuals $\hat P_{it}^\perp = 
P_{it} - \hat \gamma^P_{t,\ell}(S_{it}(\hat \phi))$ and $\hat Q_{it}^\perp = 
Q_{it} - \hat \gamma^Q_{t,\ell}(S_{it}(\hat \phi))$ for $i \in I_\ell$. Then, obtain the slope estimator as:
$$
\hat \delta_t = 
(\sum_{i=1 }^n(\hat P_{it}^\perp)^2  )^{-1}
\sum_{i=1}^n \hat P_{it}^\perp \hat Q_{it}^\perp.
$$
Then, appealing to the theoretical results in \cite{chernozhukov2018} for partially linear models, we obtain \( \sqrt{n} \)-consistency and asymptotic normality for $\delta_t$.

\begin{cor}
Work with conditions of the previous proposition. In addition, assume that $\mathrm{E}[(P_{it}^{\perp})^2]$ is bounded away from zero. Then
$
\sqrt{n} (\hat \delta_t - \delta_t) = \{\mathrm{E}[(P_{it}^{\perp})^2]\}^{-1} \frac{1}{\sqrt{n}}
\sum_{i=1}^n P^\perp_{it} e_{it} + o_p(1) \to_d  N(0, V),
$
where $V = \mathrm{E}[(P_{it}^{\perp})^2]^{-2} \mathrm{E}[(P_{it}^{\perp})^2 e^2_{it}].$ 
\end{cor}

In summary, with a sufficiently large sample size for fine-tuning and highly informative embeddings, it is plausible that the projection function \( \mathrm{E}[Y_{it} \mid S_{it}] \) can be approximated well enough to support standard asymptotic inference.

%% file: JASA_Buybox_Version/OnlineAppendix.tex

\begin{center}\Large Online Appendix \\ to \\ \textit{Adventures in Demand Analysis Using AI}  \end{center}

\section{Workflow}\label{app:workflow}

The workflow is based on the dataset $(\text{Data}_{it})_{i\in I, t \in\{0,\dots T\}}$, described in Section 2, where $I$ denotes the set of all collected product IDs ("ASIN") and $\{0,\dots T\}$ the observed time periods.

\begin{enumerate}
    \item \textbf{Dataset Split:}  
    Divide the dataset $(\text{Data}_{it})_{i \in I, t \in \{0, \dots, T\}}$ into two disjoint subsets based on product IDs: $I = I_1 \cup I_2$.
    
    \item \textbf{Embedding Fine-Tuning:}  
    Fine-tune the high-dimensional product embeddings $\tilde{E}_i$ on $I_1$ by training the neural network described in Section \ref{app:impl} using the following loss function:  
    $$
    L= \|Q_{it} - \hat{l}(X_i^{in})\|_{pred,2} \cdot \|P_{it} - \hat{m}(X_i^{in})\|_{pred,2},
    $$
    where $\|\cdot\|_{pred,2}$ is the empirical root mean square error (empirical prediction norm).
    \item \textbf{Embedding Computation:}  
    Compute high-dimensional product embeddings $\tilde{E}_i \in \mathbb{R}^{1888}$ for all product IDs $i \in I$.
    
    \item \textbf{Feature Computation:}  
    Following Algorithm \ref{alg:features}, compute $256$-dimensional embeddings $E_i$ as well as PCA-based and similarity features $PC_{i,k}$ and $CS_{i,k}$ for $k = 1, \dots, 5$. Perform this computation for all observations $(i, t)$ with $i \in I$ and $t \in \{0, \dots, T\}$.
    
    \item \textbf{Partialling-Out:}  
    For all observations $(i, t)$ where $i \in I_2$ and $t \in \{1, \dots, T\}$, partial out the effects of $(Q_{i, t-1}, P_{i, t-1}, X_{i}^{in}, X_{it}^{o})$ to obtain $\hat Q^\perp_{it}$ and $\hat P^\perp_{it}$:  
    \begin{enumerate}
        \item Use a \textbf{linear specification} (Algorithm \ref{alg:linear_diff}) controlling for either $V_i(X_i^{in}) = CS_i$ or $V_i(X_i^{in}) = E_i$.  
        \item Use a \textbf{partially linear specification} (Algorithm \ref{alg:plr_diff}) controlling for either $V_i(X_i^{in}) = CS_i$ or $V_i(X_i^{in}) = E_i$.  
    \end{enumerate}
    
    \item \textbf{Homogeneous Effect:}  
    Estimate the homogeneous effect $\hat{\alpha}$ by performing a linear regression:  
    $
    \hat Q^\perp_{it} \sim \hat P^\perp_{it},$
    clustering the standard errors at the product ID level ("ASIN").
    
    \item \textbf{Heterogeneous Effect:}  
    To estimate the heterogeneous effect $\hat{\alpha}(X_{it}^{e})$, perform a linear regression:  
    $$
    \hat Q^\perp_{it} \sim \hat P^\perp_{it} \cdot \big(1 + Q_{i, t-1} + P_{i, t-1} + CS_{i1} + CS_{i2} + CS_{i3} + CS_{i4} + CS_{i5}\big),
    $$
    clustering the standard errors at the product ID level ("ASIN").
\end{enumerate}

The PCA algorithm and cluster similarity analysis, including $k$-means, are implemented using the \texttt{scikit-learn} package \citep{scikit-learn}. Linear regression models are performed using the \texttt{statsmodels} package \citep{seabold2010statsmodels}. For the estimation of partially linear models, the \texttt{DoubleML} package \citep{bach2022doubleml, bach2024doubleml} is used, with boosting algorithms implemented via the \texttt{Lightgbm} package \citep{ke2017lightgbm}. To simplify computations in high-dimensional settings, the embeddings are projected to a lower-dimensional space using Gaussian Random Projection, which preserves distances as described in \citep{johnson1984extensions}.

\begin{algorithm}[H] \label{alg:features}
\caption{Generating PCA and Similarity Features}
\KwIn{Embeddings ${E}_i\in \mathbb{R}^d$ for $i \in I$.}
\KwOut{Projected and Normalized Embeddings $X^e_i\in \mathbb{R}^{256}$, Cluster Centroids $CS_i$, and Principal Components $PC_i$.}

Project embeddings ${E}_i$ onto a lower-dimensional space $\bar E_i \gets {E}_i^TG$, where $G\in \mathbb{R}^{d\times 256}$ is a random matrix with i.i.d. N(0,1) entries\;

Center and normalize embeddings: $X_i^{e} \gets \frac{\bar E_i - 
\frac{1}{n} \sum_{i \in I} \bar E_i}{\|\bar E_i - \frac{1}{n} \sum_{i \in I} \bar E_i\|_2}$\;

\nl \textbf{Step 1: PCA Features}\;
\begin{itemize}
    \item Run a PCA algorithm on $(X_i^{e})_{i \in I}$\;
    \item Project each $X_i$ onto the first $k$-th principal component $PC_{ik}:= PC_k(X_i^{e})$ for $k = 1, \dots, K$\;
\end{itemize}

\nl \textbf{Step 2: Similarity Features}\;
\begin{itemize}
    \item Use $k$-means clustering with euclidean distance on $(X_i^{e})_{i \in I}$ to compute $K$ centroids $c_k$\;
    \item Compute cosine similarities $CS_{ik}:= CS_k(X_i^{e}) = \frac{c_k^T X_i^{e}}{\|c_k\|_2 \|X_i^{e}\|_2}$ for $k = 1, \dots, K$\;
\end{itemize}

\nl Return $X^e_i$, $CS_i = (CS_{i1},\dots, CS_{iK})$ and $PC_i = (PC_{i1},\dots, PC_{iK})$\;
\end{algorithm}

\begin{algorithm}[H] \label{alg:linear_diff}
\caption{Partialling-out: Linear Specification}
\KwIn{Data $(Q_{it}, P_{it}, Q_{i,t-1}, P_{i,t-1}, V_i, X_{it}^{o})$ for $i \in I_2$, $t \in \{1,\dots T\}$}
\KwOut{Partialling out values $\hat Q^\perp_{it}$, $\hat P^\perp_{it}$}

Run ordinary least squares \begin{align*}
    Q_{it} &\sim Q_{i,t-1} + P_{i,t-1} + V_i + X_{it}^{o}, \quad \quad 
    P_{it} \sim Q_{i,t-1} + P_{i,t-1} + V_i + X_{it}^{o},
\end{align*}
and keep predicted values  $\hat Q_{it}$ and $\hat P_{it}$.

Output the residuals 
$     \hat Q^\perp_{it} := Q_{it} - \hat Q_{it},$ and $\hat P^\perp_{it} := P_{it} - \hat P_{it}.
 $

\end{algorithm}

\begin{algorithm}[H] \label{alg:plr_diff}
\caption{Partialling-out: Partially Linear Specification}
\KwIn{Data $(Q_{it}, P_{it}, Q_{i,t-1}, P_{i,t-1}, V_i, X_{it}^{o})$ for $i \in I_2$, $t \in \{1,\dots T\}$}
\KwOut{Partialling out values $\hat Q^\perp_{it}$, $\hat P^\perp_{it}$}

Employ $5$-fold cross-fitting to estimate corresponding non-linear regression functions
\begin{align*}
    l_0(Q_{i,t-1}, P_{i,t-1}, V_i, X_{it}^{o}) &:= \mathrm{E}[Q_{it}|Q_{i,t-1}, P_{i,t-1}, V_i, X_{it}^{o}] \\
    m_0(Q_{i,t-1}, P_{i,t-1}, V_i, X_{it}^{o}) &:= \mathrm{E}[P_{it}|Q_{i,t-1}, P_{i,t-1}, V_i, X_{it}^{o}]
\end{align*}
via boosted trees.

Compute and report the residuals  \begin{align*}
     \hat Q^\perp_{it} &:= Q_{it} -  \hat{l}_0(Q_{i,t-1}, P_{i,t-1}, V_i, X_{it}^{o})\\
     \hat P^\perp_{it} &:= P_{it} - \hat{m}_0(Q_{i,t-1}, P_{i,t-1}, V_i, X_{it}^{o})
 \end{align*}
\end{algorithm}

\input{JASA_Buybox_Version/Appendix_substitute_prices}

\clearpage

\input{JASA_Buybox_Version/Appendix_shared_weights}

\clearpage

\input{JASA_Buybox_Version/Appendix_sensitivity}

\clearpage

\input{JASA_Buybox_Version/Appendix_clothes}

\clearpage

\section{Neural Network Architecture}\label{app:impl}

\begin{figure}[ht]
    \centering
    \includegraphics[width=\linewidth]{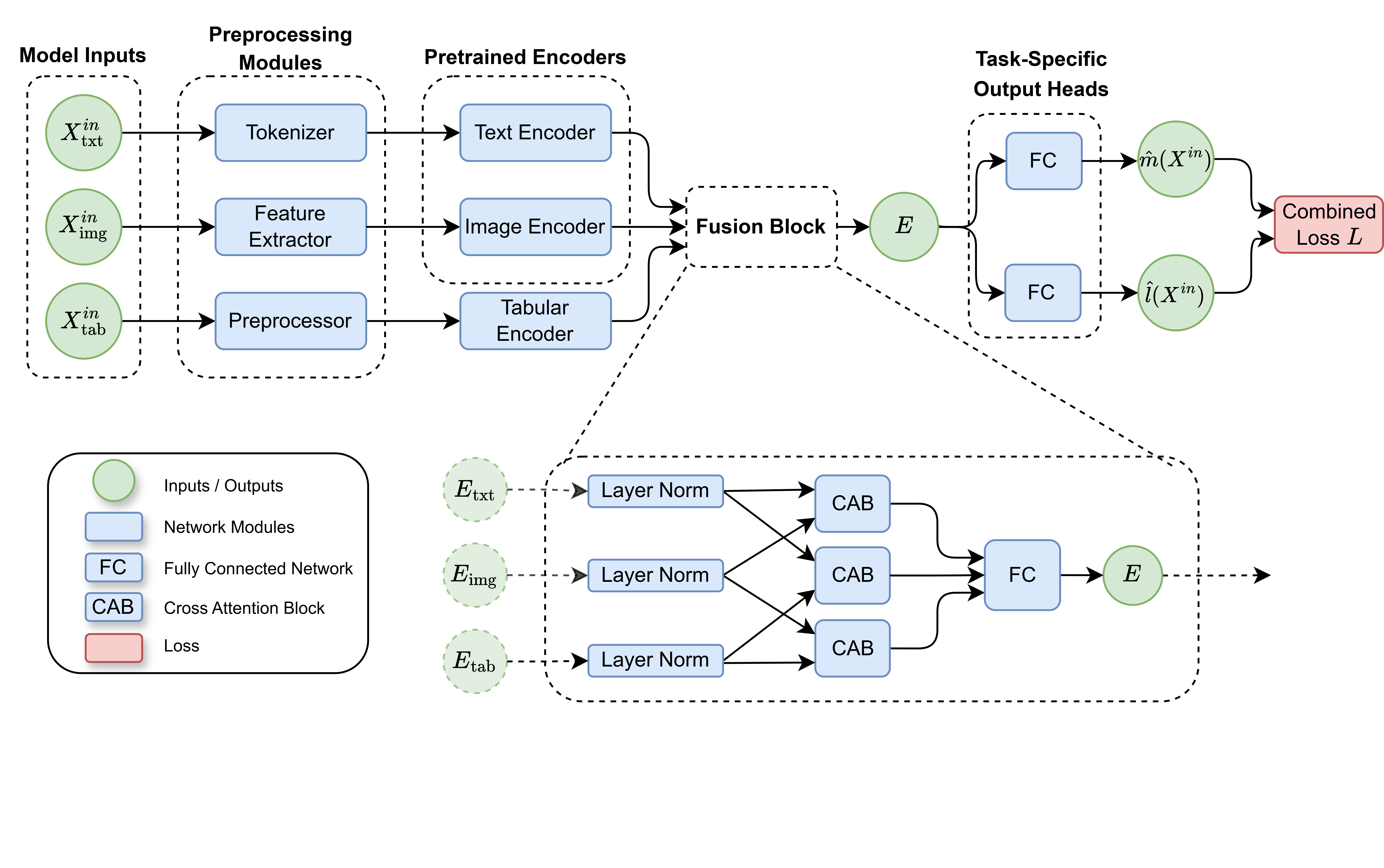}
    \caption{Proposed model architecture}
    \label{fig:model_arch}
\end{figure}

To model demand based on multimodal data inputs, we develop and implement an architecture presented in Figure \ref{fig:model_arch}. As our interest is not only in predicting the quantity demanded $Q_{it}$, but also on valid statistical inference on elasticity parameters, our architecture has been adapted to the double/debiased machine learning framework \citep{chernozhukov2018}. The later is based on an orthogonal moment condition to account for typical machine-learning induced biased and, hence, requires to additionally perform predictions on the price variable $P_{it}$, see \cite{klaassen2024doublemldeep} for more details. Each of these transformer blocks will output a dense embedding, which are later being concatenated to the multimodal embedding $E_{i}$. 

The following model components were used for implementation: The SAINT model \citep{somepalli2021saintimprovedneuralnetworks} implemented in the \texttt{pytorch-widedeep} package \citep{Zaurin_pytorch-widedeep_A_flexible_2023} has been used as the tabular encoder, RoBERTa model being pretrained on a Twitter Dataset \citep{loureiro2022timelms} as the text encoder, and the BEiT model \citep{bao2022beit} as the image encoder.




%% file: JASA_Buybox_Version/Appendix_substitute_prices.tex
\section{Robustness Check: Including Substitute Prices}

Substitute prices can have a strong influence on the price elasticity of products. As a robustness check of our model specification, we are including the quantity-weighted price of the $5$ closest substitutes (with respect to our estimated embedding) as an additional control variable. 

\begin{figure}[h]
    \centering
    \includegraphics[width=0.6\linewidth]{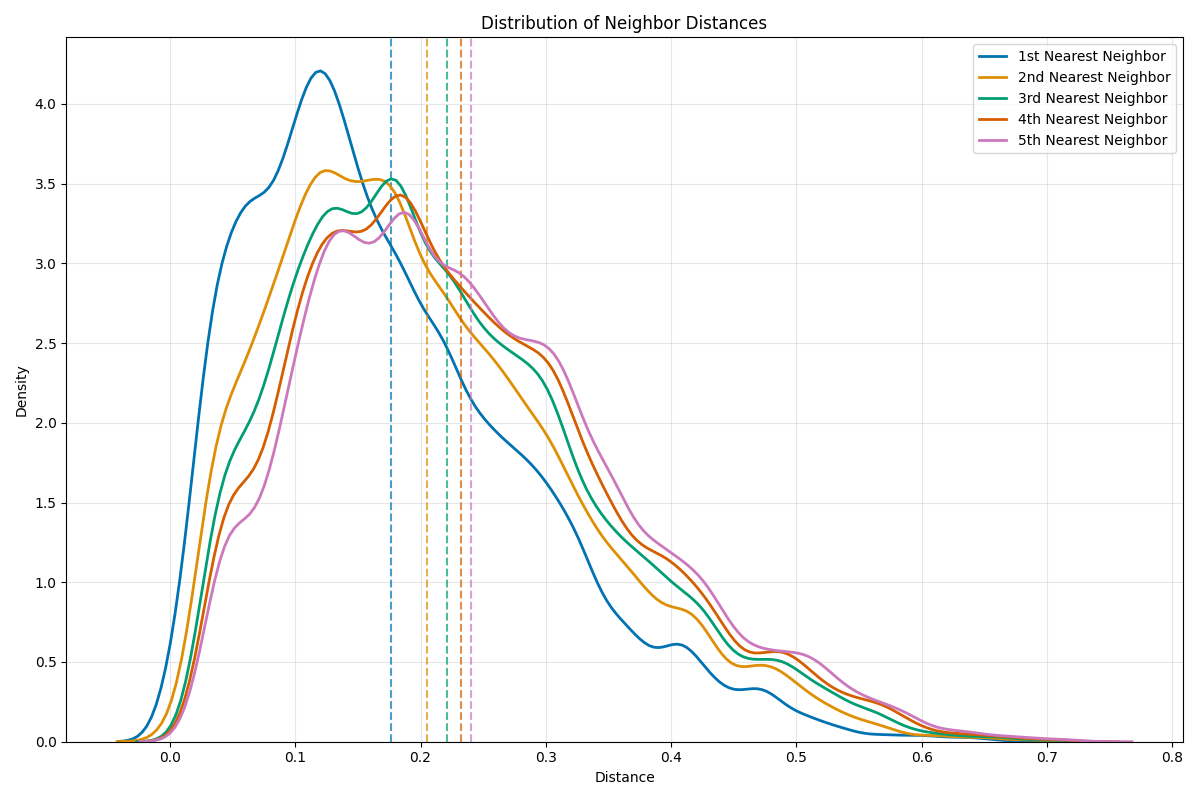}
    \caption{Distribution of Neighbour Distances}
    \label{fig:neighb_dist_sp}
\end{figure}

For each product $i$, we collect the $5$ closest neighbors $N(i)$ with respect to cosine distance, i.e. the $5$ products with the highest cosine similarity (see Figure \ref{fig:neighb_dist_sp}).

The quantity-weighted price is defined as
\begin{align*}
    X^{sub}_{it}:=\sum_{j\in N(i)} w_{ijt}P_{jt},
\end{align*}
where the weights are determined via the quantity signal
\begin{align*}
    w_{ijt}:= \frac{\exp(Q_{j, t-1})}{\sum_{m\in N(i)} \exp(Q_{m, t-1})}.
\end{align*}

\begin{figure}[h]
    \centering
    \includegraphics[width=0.6\linewidth]{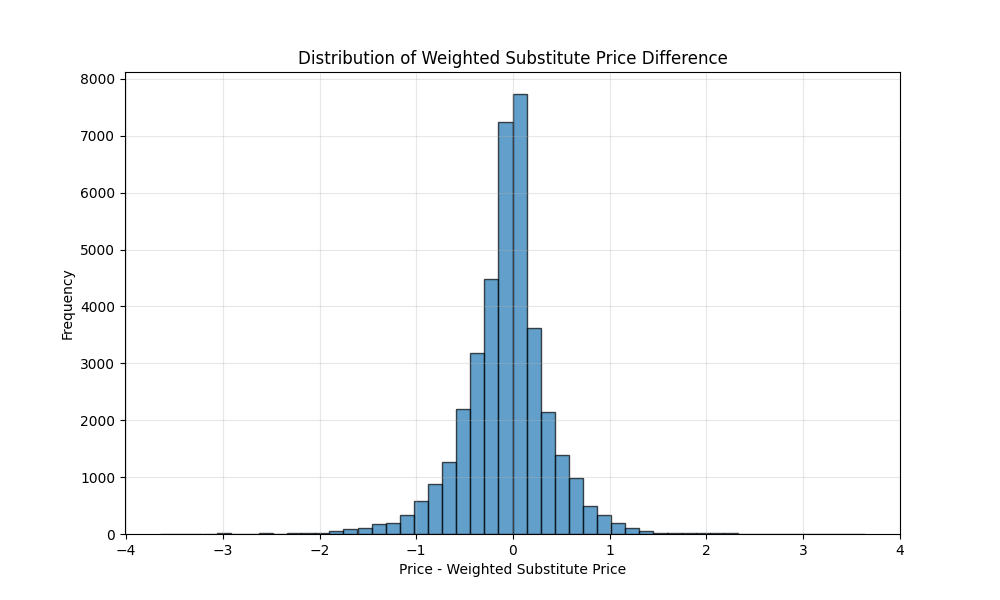}
    \caption{Distribution of $P_{it} - X_{it}^{sub}$}
    \label{fig:subdiff}
\end{figure}

The difference $P_{it} - X_{it}^{sub}$ is shown in Figure \ref{fig:subdiff}

Remark that we only focus on neighbors, where previous quantities are measured, such that we can compute the weights $w_{j,t}$.
In the following, we repeated the analysis of original analysis, but added the quantity-weighted price as additional control
\begin{align*}
    \tilde{X}^o_{it}:= (X^o_{it}, X^{sub}_{it}).
\end{align*}

All results are reported as in the main paper, differences are calculated as $\Delta = v_{sub} - v_{main}$ for all values $v$.

\subsection{Homogeneous Elasticity Model}

\begin{table}[h]
\centering
\caption{Estimated price effects based on the partially linear dynamic model (including substitute prices).}
\label{tab:dynamic_effect_analysis_sp}
\resizebox{1\textwidth}{!}{
\begin{tabular}{lcccccc}
\hline\hline
  Specification of Control Function (State $S_{t}$)             & {coef} & {std err} & {t} & {P-val.} & {[5.0\%} & {95.0\%]} \\ \hline
{I-1. Linear ($P_{t-1}$, $Q_{t-1}$)}                        & -0.690 ($\Delta=0$) & 0.040 ($\Delta=0$) & -17.248 ($\Delta=0$) & 0.000 ($\Delta=0$) & -0.756 ($\Delta=0$) & -0.624 ($\Delta=0$)\\ 
{I-1. Linear ($P_{t-1}$, $Q_{t-1}$, $X^{e}, \tilde{X}^o_{t}$)}      & -0.715 ($\Delta=-0.003$) & 0.039 ($\Delta=0$) & -18.413 ($\Delta=-0.049$) & 0.000 ($\Delta=0$) & -0.779 ($\Delta=-0.003$) & -0.651 ($\Delta=-0.002$)\\ 
{I-1. Linear ($P_{t-1}$, $Q_{t-1}$, $X^{sim}, \tilde{X}^o_{t}$)}    & -0.725 ($\Delta=-0.002$) & 0.039 ($\Delta=0$) & -18.725 ($\Delta=0$) & 0.000 ($\Delta=0$) & -0.789 ($\Delta=-0.003$) & -0.662 ($\Delta=-0.003$) \\
{I-2. Linear with Interactions ($P_{t-1}$, $Q_{t-1}$, $X^{sim}, \tilde{X}^o_{t}$)}    & -0.729 ($\Delta=-0.002$) & 0.039 ($\Delta=0$) & -18.698 ($\Delta=0$) & 0.000 ($\Delta=0$)  & -0.793 ($\Delta=-0.002$) & -0.665 ($\Delta=-0.003$)\\
{I-3. Boosted Trees ($P_{t-1}$, $Q_{t-1}$, $X^{e}, \tilde{X}^o_{t}$)}    &  -0.699 ($\Delta=-0.076$) & 0.048 ($\Delta=-0.001$) & -14.467 ($\Delta=-1.835$) & 0.000 ($\Delta=0$) & -0.778 ($\Delta=-0.074$) & -0.619 ($\Delta=-0.077$)\\
{I-3. Boosted Trees ($P_{t-1}$, $Q_{t-1}$, $X^{sim}, \tilde{X}^o_{t}$)}    & -0.696 ($\Delta=-0.005$) & 0.042 ($\Delta=+0.001$) & -16.716 ($\Delta=+0.335$) & 0.000 ($\Delta=0$) & -0.765 ($\Delta=-0.008$) & -0.628 ($\Delta=-0.004$)\\
\hline\hline
\end{tabular}}
  \caption*{\footnotesize Note: Standard errors are clustered at the product level. The LR models are estimated using OLS. The PLR model is estimated using DML with cross-fitted boosted trees.}
\end{table}

\break

\subsection{Heterogeneous Elasticity Model}

\begin{table}[h]
\small
\centering
\caption{Inference on the price effect modifiers with nonlinear (Boosted Trees) state control (Model II-3, including substitute prices). Results for Models II-1 and II-2 are similar.}
\label{tab:dynamic_effect_analysis_cate_PLR_sp}
\resizebox{1\textwidth}{!}{
\begin{tabular}{lcccccc}
\hline\hline
   Modifier              & {coef} & {std err} & {t} & {P-val.} & {[5.0\%} & {95.0\%]} \\ \hline
{Centercept} & -0.739 ($\Delta=-0.003$) & 0.046 ($\Delta=0$) & -15.952 ($\Delta=-0.092$) & 0.000 ($\Delta=0$) & -0.815 ($\Delta=-0.003$) & -0.663 ($\Delta=-0.004$) \\
{Lagged Quantity, $Q_{t-1}$} & -0.326 ($\Delta=-0.001$) & 0.096 ($\Delta=+0.004$) & -3.385 ($\Delta=+0.130$) & 0.001 ($\Delta=+0.001$) & -0.484 ($\Delta=-0.007$) & -0.168 ($\Delta=+0.005$) \\
{Lagged Price, $P_{t-1}$} & -0.052 ($\Delta=-0.003$) & 0.038 ($\Delta=+0.003$) & -1.392 ($\Delta=+0.009$) & 0.164 ($\Delta=+0.003$) & -0.114 ($\Delta=-0.008$) & 0.009 ($\Delta=+0.001$) \\
{Cluster Similarity $0$} & -1.607 ($\Delta=+0.705$) & 1.544 ($\Delta=+0.020$) & -1.041 ($\Delta=+0.476$) & 0.298 ($\Delta=+0.169$) & -4.146 ($\Delta=+0.673$) & 0.932 ($\Delta=+0.737$) \\
{Cluster Similarity $1$} & -7.490 ($\Delta=-0.125$) & 1.836 ($\Delta=+0.005$) & -4.079 ($\Delta=-0.056$) & 0.000 ($\Delta=0$) & -10.511 ($\Delta=-0.134$) & -4.470 ($\Delta=-0.116$) \\
{Cluster Similarity $2$} & -8.477 ($\Delta=-1.692$) & 3.571 ($\Delta=+0.013$) & -2.374 ($\Delta=-0.467$) & 0.018 ($\Delta=-0.038$) & -14.350 ($\Delta=-1.713$) & -2.604 ($\Delta=-1.671$) \\
{Cluster Similarity $3$} & -10.074 ($\Delta=-1.819$) & 3.993 ($\Delta=+0.008$) & -2.523 ($\Delta=-0.451$) & 0.012 ($\Delta=-0.026$) & -16.642 ($\Delta=-1.832$) & -3.507 ($\Delta=-1.807$) \\
{Cluster Similarity $4$} & -4.895 ($\Delta=-0.336$) & 1.294 ($\Delta=-0.009$) & -3.783 ($\Delta=-0.283$) & 0.000 ($\Delta=0$) & -7.023 ($\Delta=-0.321$) & -2.766 ($\Delta=-0.349$) \\
\hline\hline
\end{tabular}}
\caption*{\footnotesize Note: Standard errors are clustered at the product level. The LR model is estimated using OLS. The PLR model is estimated using DML with cross-fitted boosted trees. Lagged quantities and prices are centered and rescaled to have unit variance across \((i,t)\).  The coefficient on the centercept represents the average effect. }
\end{table}

\begin{figure}[ht]
    \centering
    \includegraphics[width=0.8\linewidth]{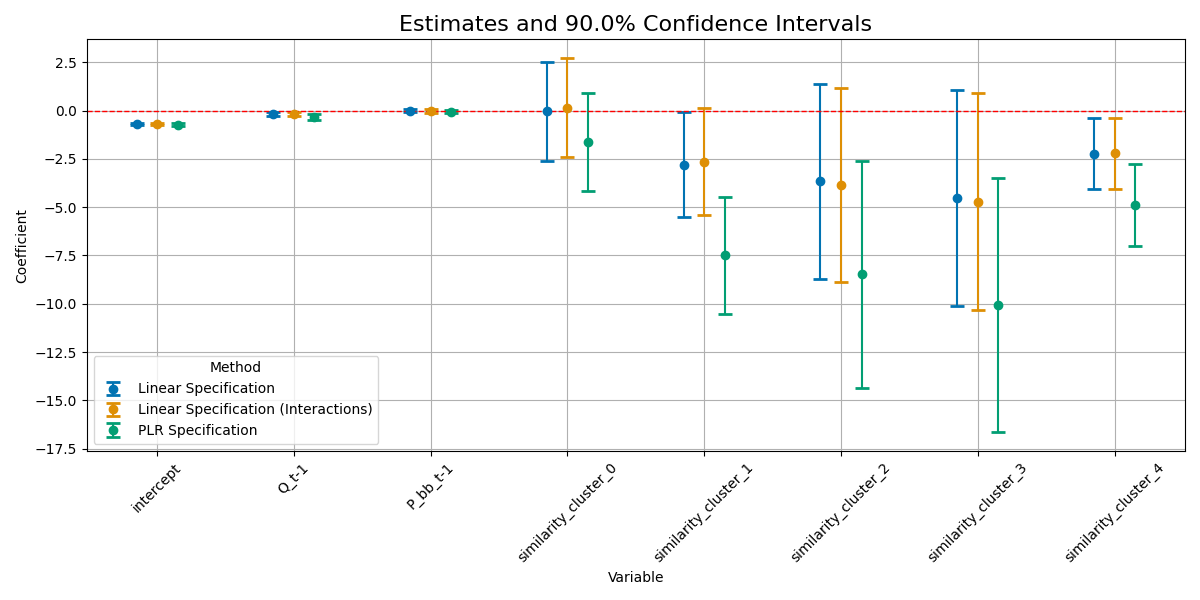}
    \caption{Parameters of the Estimated Elasticity Function and 90\% Confidence Intervals (with substitute prices as control variable)}
    \label{fig:cate_sub}
\end{figure}

\begin{table}[h]\footnotesize
    \centering
    \caption{$p$-values for $\chi^2$-test of joint significance of price effect modifiers (including substitute prices as control variable)}
    \label{tab:cluster_significance_dynamic_sp}
    \begin{tabular}{@{}lcc@{}}
        \hline\hline
        Model & All Modifiers  &   Similarities Only\\ \midrule
        II.1 Linear Specification & 0.000 ($\Delta=0$) & 0.033 ($\Delta=+0.003$) \\
        II.2 Linear Specification (Interactions) & 0.000 ($\Delta=0$) & 0.056 ($\Delta=+0.002$) \\
        II.3 PLR Specification (Boosted Trees) & 0.000 ($\Delta=0$) & 0.001 ($\Delta=0$) \\ \hline\hline
    \end{tabular}
    \caption*{\footnotesize Note: Standard errors are clustered at the product level. LR is estimated using OLS. The PLR model is estimated using DML with cross-fitted boosted trees (sample size: $38,041$).}
\end{table}

\begin{figure}[ht]
    \centering
    \includegraphics[width=0.8\linewidth]{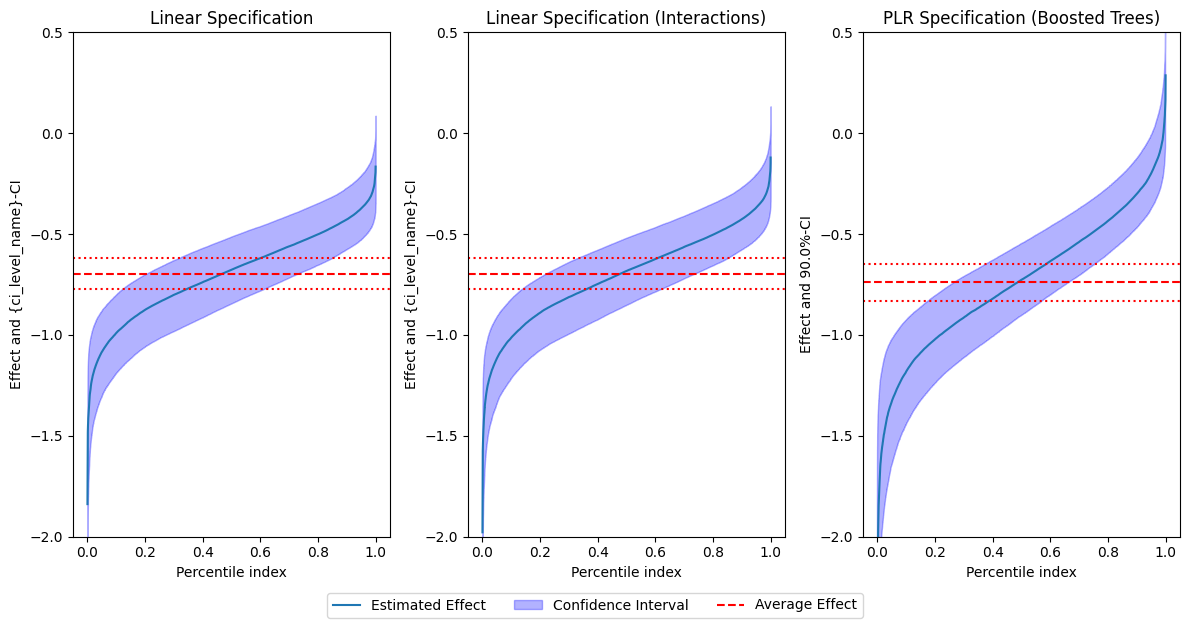}
    \caption{Sorted Elasticity as a Function of Effect Modifiers and 90\% Pointwise Confidence Bands (including substitute prices as control variable)}
    \label{fig:sorted_sub}
\end{figure}

Overall the results are very similar to the main analysis.

%% file: JASA_Buybox_Version/Appendix_shared_weights.tex
\section{Dimensionality Reduction via Neural Networks}

In the main text, we employed a Johnson-Lindenstrauss projection to compress the high-dimensional concatenated embeddings $E_i$ into a lower-dimensional vector $\tilde{E}_i$. While that approach offers theoretical guarantees regarding the preservation of pairwise distances without optimization, one might naturally ask: \textit{Why not perform dimensionality reduction directly to a low-dimensional embedding (e.g., 5 dimensions) using a neural network, bypassing the projection-based approach?}

In this section, we describe an approach that replaces the random projection with a learned non-linear compression map. This method aligns with the principle of causal fine-tuning discussed in Remark 3, as it forces the dimensionality reduction to retain only the information most relevant to the prediction of price and quantity signals.
Instead of projecting the embeddings $E_i$ onto a random subspace, we utilize a multi-task neural network architecture to learn a compact representation. Let $X_i^{in}$ denote the raw multimodal input. We define a feed-forward neural network $f_{\theta}(\cdot)$ parametrized by weights $\theta$, which maps the input to a low-dimensional bottleneck layer $Z_i \in \mathbb{R}^d$ (where $d=5$ in this experimental setup).

The architecture proceeds as follows:

\begin{enumerate}
    \item Shared Representation (Encoder): The input passes through several hidden layers with non-linear activations (e.g., ReLU), culminating in a final hidden layer of dimension $d=5$. We denote this bottleneck representation as the learned embedding:
    $$Z_i = f_{\theta}(X_i^{in})$$
    \item Linear Readout (Decoder): To ensure $Z_i$ captures information relevant to our economic variables, we employ a final output layer with an ELU activation function. Crucially, this output layer is used to predict the vector of targets $Y_{it} = [Q_{it}, P_{it}]^{\top}$. Here, $W_{out}$ and $b_{out}$ represent the weights and biases of the linear prediction head.
\end{enumerate}

In this approach the parameters $\theta$ of the encoder $f_{\theta}$ are shared across the prediction tasks for both Price ($P_{it}$) and Quantity ($Q_{it}$). By optimizing the network to minimize the joint loss function over both signals, we force the 5-dimensional embedding $Z_i$ to become a sufficient summary statistic for both demand and pricing mechanisms.
Unlike the unsupervised Gaussian random projection, this approach is supervised. The resulting embeddings are not guaranteed to preserve all geometric distances from the original space, but are optimized to preserve predictive variations relevant to $Q_{it}$ and $P_{it}$. The optimized representations for $Q_{it}$ and $P_{it}$ do not automatically imply optimization for elasticity heterogeneity, as can be seen in Table \ref{tab:cluster_significance_dynamic_shared_weights}.

\subsection{Homogeneous Elasticity Model}

\begin{table}[h]
\centering
\caption{Estimated price effects based on the partially linear dynamic model.}
\label{tab:dynamic_effect_analysis_shared_weights}
\resizebox{1\textwidth}{!}{
\begin{tabular}{lcccccc}
\hline\hline
  Specification of Control Function (State $S_{t}$)             & {coef} & {std err} & {t} & {P-val.} & {[5.0\%} & {95.0\%]} \\ \hline
{I-1. Linear ($P_{t-1}$, $Q_{t-1}$)}                        & -0.690 & 0.040 & -17.248 & 0.000 & -0.756 & -0.624 \\ 
{I-1. Linear ($P_{t-1}$, $Q_{t-1}$, $Z, X^o_{t}$)} & -0.746 & 0.039 & -18.413 & 0.000 & -0.77 & -0.651 \\ 
{I-2. Linear with Interactions ($P_{t-1}$, $Q_{t-1}$, $Z, X^o_{t}$)}    & -0.750 & 0.039 & -19.152 & 0.000 &  -0.814 &  -0.686   \\
{I-3. Boosted Trees ($P_{t-1}$, $Q_{t-1}$, $Z, X^o_{t}$)}    &  -0.709 & 0.042 & -17.036 & 0.000  & -0.777 &-0.640\\
\hline\hline
\end{tabular}}
  \caption*{\footnotesize Note: Standard errors are clustered at the product level. The LR models are estimated using OLS. The PLR model is estimated using DML with cross-fitted boosted trees.}
\end{table}

\break

\subsection{Heterogeneous Elasticity Model}

\begin{table}[h]
\small
\centering
\caption{Inference on the price effect modifiers (Notebook Model II-3) with embedding features (PLR with boosted trees)}
\label{tab:dynamic_effect_analysis_cate_shared_weights}
\begin{tabular}{lcccccc}
\hline\hline
Modifier & {coef} & {std err} & {t} & {P-val.} & {[5.0\%} & {95.0\%]} \\ \hline
Centercept & -0.727 & 0.040 & -18.349 & 0.000 & -0.792 & -0.662 \\
Lagged Quantity $Q_{t-1}$ & -0.174 & 0.054 & -3.225 & 0.001 & -0.263 & -0.085 \\
Lagged Price $P_{t-1}$ & 0.012 & 0.031 & 0.398 & 0.691 & -0.039 & 0.064 \\
$Z_1$ & 2.166 & 2.176 & 0.995 & 0.320 & -1.414 & 5.745 \\
$Z_2$ & -2.272 & 3.002 & -0.757 & 0.449 & -7.209 & 2.666 \\
$Z_3$ & -0.126 & 0.100 & -1.259 & 0.208 & -0.290 & 0.039 \\
$Z_4$ & 0.073 & 9.279 & 0.008 & 0.994 & -15.189 & 15.335 \\
$Z_5$ & -0.703 & 3.669 & -0.192 & 0.848 & -6.737 & 5.331 \\
\hline\hline
\end{tabular}
\end{table}

\begin{figure}[ht]
    \centering
    \includegraphics[width=0.8\linewidth]{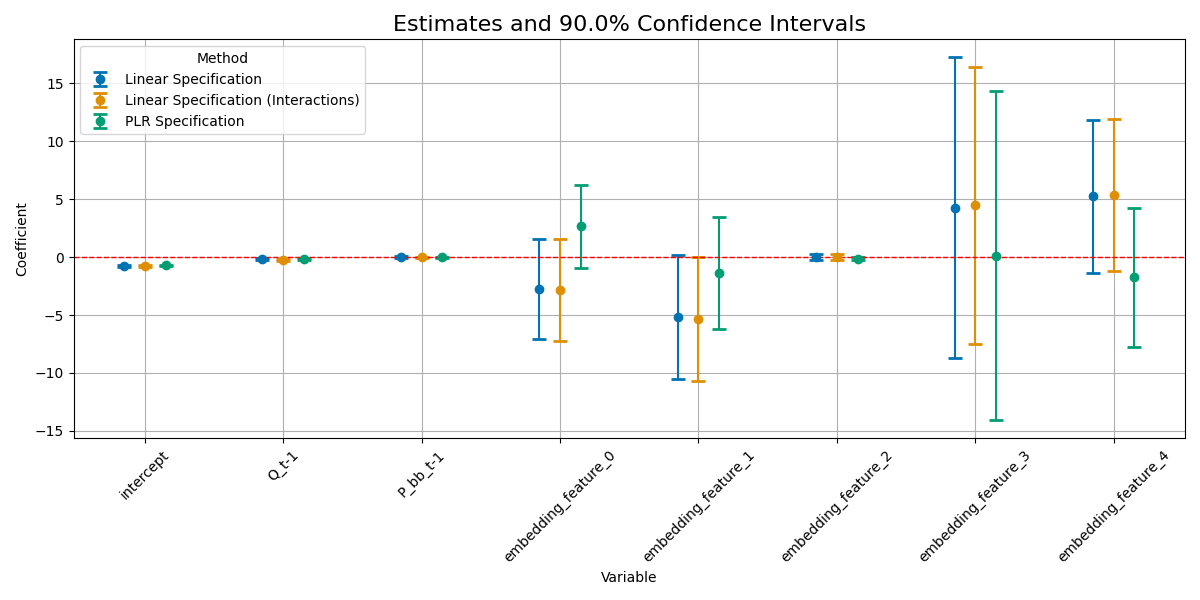}
    \caption{Parameters of the Estimated Elasticity Function and 90\% Confidence Intervals}
    \label{fig:cate_sw}
\end{figure}

\begin{table}[ht]\footnotesize
\centering
\caption{$p$-values for $\chi^2$-test of joint significance}
\label{tab:cluster_significance_dynamic_shared_weights}
\begin{tabular}{@{}lcc@{}}
\hline\hline
Model & All Modifiers & $Z$ Only \\ \midrule
II.1 Linear Specification & 0.000 ($\Delta = 0.000$) & 0.526 ($\Delta = +0.496$) \\
II.2 Linear Specification (Interactions) & 0.000 ($\Delta = 0.000$) & 0.443 ($\Delta = +0.389$) \\
II.3 PLR Specification (Boosted Trees) & 0.000 ($\Delta = 0.000$) & 0.007 ($\Delta = +0.006$) \\ \hline\hline
\end{tabular}
  \caption*{\footnotesize Note: Standard errors are clustered at the product level. LR is estimated using OLS. The PLR model is estimated using DML with cross-fitted boosted trees (sample size: $38,041$).}
\end{table}

\begin{figure}[ht]
    \centering
    \includegraphics[width=0.8\linewidth]{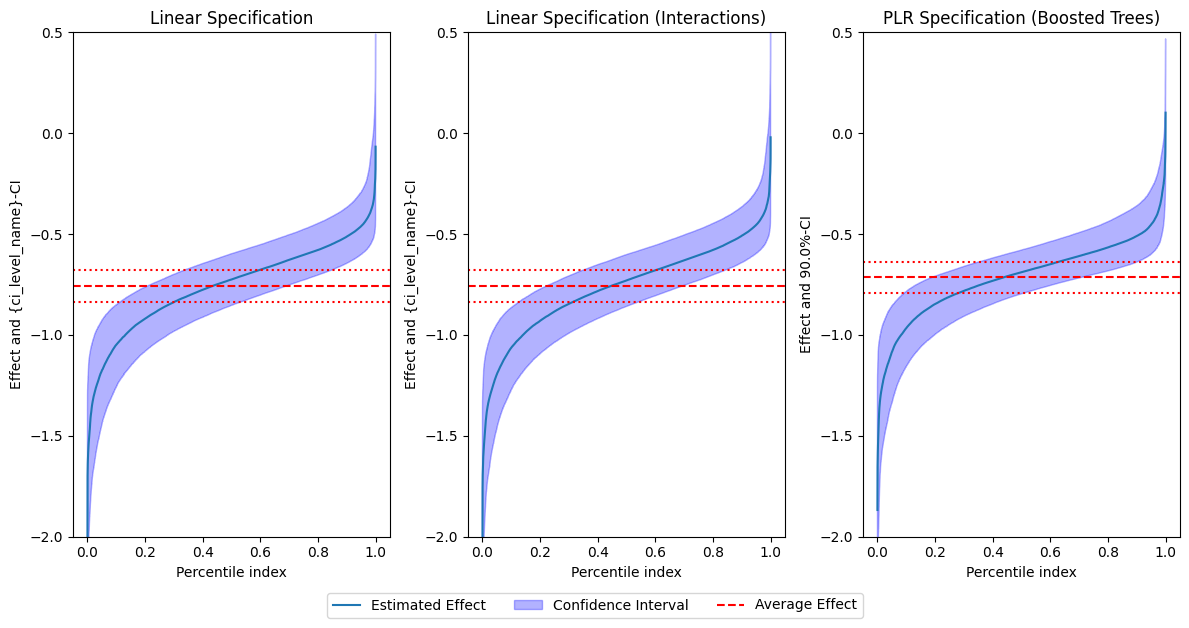}
    \caption{Sorted Elasticity as a Function of Effect Modifiers and 90\% Pointwise Confidence Bands}
    \label{fig:sorted_sw}
\end{figure}

%% file: JASA_Buybox_Version/Appendix_sensitivity.tex
\section{Robustness Check: Sensitivity Analysis}

This section includes a sensitivity analysis with respect to omitted confounding as proposed in \citet{chernozhukov2021long}.
The approach quantifies how strong an unobserved confounder would need to be, in terms of explanatory power, to affect the estimated treatment effect.

The sensitivity parameters \(R^2_Y\) and \(R^2_D\) are nonparametric partial \(R^2\) measures that capture the proportion of residual variation explained by a hypothetical omitted confounder. Specifically, \(R^2_Y\) denotes the share of unexplained variation in the outcome, i.e. $Q_{it}$, that would be accounted for by the confounder after conditioning on the observed state $S_{it}$, while \(R^2_D\) represents the corresponding share of residual variation in the treatment variable, i.e. $P_{it}$.

The parameter \(\rho\) governs the correlation between the confounder’s effects on the outcome and on the treatment. Following \citet{chernozhukov2021long}, \(\rho = 1\) corresponds to a worst-case scenario in which the confounder affects the outcome and the treatment in perfectly aligned directions, thereby maximizing potential bias. Lower values of \(\rho\) reflect weaker alignment between these effects and hence less adversarial confounding structures.

By varying \(\rho\) and the magnitude of \((R^2_Y, R^2_D)\), the sensitivity analysis traces how the estimated treatment effect and its associated uncertainty evolve under increasingly strong and correlated omitted confounding. The bounds reported in Tables~\ref{tab:sensitivity_rho_10}--\ref{tab:sensitivity_rho_02} refer to the price effects with nonlinear (boosted-trees) state control (Model I-3).
In the following tables, \(\theta_{-}\) and \(\theta_{+}\) denote the lower and upper bounds of the point estimate implied by a given sensitivity scenario. Accounting additionally for sampling uncertainty yields the corresponding one-sided lower and upper confidence bounds reported in the tables.

Following the benchmarking approach of Appendix D in \citet{chernozhukov2021long}, leaving out all other tabular control variables $X_t^o$ (including such important variables  as the average reviews and their number) from the model implies a comparable confounding scenario\footnote{Following the benchmarking process yields $\tilde R^2_Y = 0.2008$, $\tilde R^2_D = 0.0499$ and $\tilde \rho=0.1503$ from the omission of $X_t^o$, which is equivalent (generates the same bias bound) as the scenario with the adversarial confounding $\rho =1$ and the pair $R^2_Y = R^2_D$ set to  $|\tilde \rho|\sqrt{\tilde R^2_Y\frac{1-\tilde \tilde R^2_D}{R^2_D}} =0.0154$. } of $R^2_Y = R^2_D = 0.0154$ for a correlation parameter $\rho=1.0$. This corresponds to confounding between the mild and medium scenarios below with $\rho= 1.0$.

\begin{table}[h]    
\centering
\caption{Sensitivity analysis for $\rho=1.0$: }
\label{tab:sensitivity_rho_10}
\begin{tabular}{lcccccc}
\hline\hline    
Scenario ($R^2_Y$, $R^2_D$) & $[ {10.0}\% $ & $\theta_-$ & $\theta$ & $\theta_+$ & ${90.0}\% ]$ \\
\midrule
mild $(.01, .01)$ & -0.802 & -0.741 & -0.697 & -0.654 & -0.591 \\
medium $(.025, .025)$ & -0.868 & -0.807 & -0.697 & -0.587 & -0.524 \\
strong $(.05, .05)$ & -0.979 & -0.919 & -0.697 & -0.475 & -0.409 \\
\hline\hline
\end{tabular}
\end{table}

\begin{table}[h]
\centering
\caption{Sensitivity analysis for $\rho=0.5$: }
\label{tab:sensitivity_rho_05}
\begin{tabular}{lcccccc}
\hline\hline
Scenario ($R^2_Y$, $R^2_D$) & $[ {10.0}\% $ & $\theta_-$ & $\theta$ & $\theta_+$ & ${90.0}\% ]$ \\
\midrule
mild $(.01, .01)$ & -0.781 & -0.719 & -0.697 & -0.675 & -0.613 \\
medium $(.025, .025)$ & -0.813 & -0.752 & -0.697 & -0.642 & -0.579 \\
strong $(.05, .05)$ & -0.869 & -0.808 & -0.697 & -0.586 & -0.522 \\
\hline\hline
\end{tabular}
\end{table}

\begin{table}[h]
\centering
\caption{Sensitivity analysis for $\rho=0.2$: }
\label{tab:sensitivity_rho_02}
\begin{tabular}{lcccccc}
\hline\hline
Scenario ($R^2_Y$, $R^2_D$) & $[ {10.0}\% $ & $\theta_-$ & $\theta$ & $\theta_+$ & ${90.0}\% ]$ \\
\midrule
mild $(.01, .01)$ & -0.768 & -0.706 & -0.697 & -0.688 & -0.626 \\
medium $(.025, .025)$ & -0.781 & -0.719 & -0.697 & -0.675 & -0.613 \\
strong $(.05, .05)$ & -0.803 & -0.742 & -0.697 & -0.653 & -0.590 \\
\hline\hline
\end{tabular}
\end{table}

%% file: JASA_Buybox_Version/Appendix_clothes.tex
\section{Replication with another product category}

A new data set for the category \textit{Clothing, Shoes \& Jewelry - Women} (browse node 7147440011) was created in the same way as the Toys data set in the main analysis.

\subsection{Qualitative Assessment}

\begin{figure}[H]
    \centering
        \centering
        \includegraphics[width=0.6\linewidth]{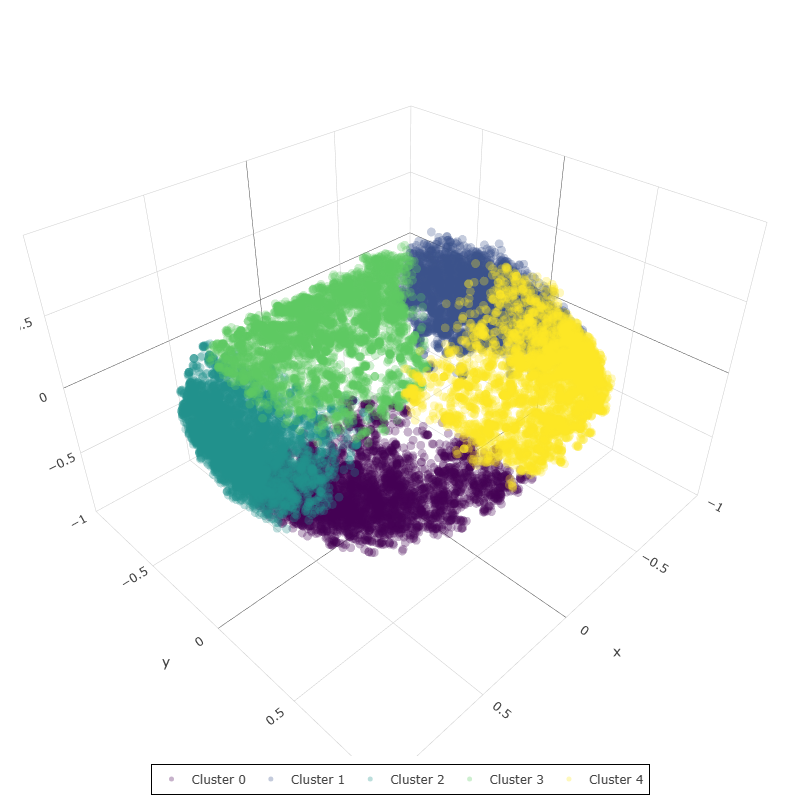}
        \caption{3d representation of product embeddings and five clusters (Clothes dataset)}
        \label{fig:cluster_pca_img_3d_comb_cloth}
\end{figure}


\clearpage

\subsection{Quantitative Assessment}

\begin{table}[ht]\small
\centering
\caption{Test $R^2$ scores for predicting quantity and price signals (Clothes dataset).}
\label{tab:r2_cloth}
\begin{tabular}{lcc|cc}
\hline\hline
{Method [features] \textbackslash Target }  & {$Q_{it}$} & {$P_{it}$} & {$\Delta Q_{it}$} & {$\Delta P_{it}$} \\ \hline
{Linear Reg  [all tabular]} & 2.94\% & 7.25\% & 2.58\% & 0.26\%\\
{Boosted Trees Reg [all tabular] } & 23.62\% & 0.72\% & 0.00\% & 0.00\%\\
{Deep Learning Reg [image and text; invariant tabular]} & 14.92\% &	66.44\% & 0.00\% & 0.00\%\\
{Deep Learning Reg [image and text; all tabular]} & 33.28\% & 63.40\% & 0.00\% & 0.00\% \\
\hline\hline
\end{tabular}
\vspace{2mm}

\caption*{ \footnotesize Note: All models are trained on a training set and scores are evaluated on a test set. Predictions use lagged values of time-varying controls. Negative $R^2$ values are set to $0$.}
\end{table}



\begin{table}[h!]\small
\centering
\caption{Test $R^2$ Scores for ML methods using DL-based PCAs and similarities together with tabular controls (Clothes dataset).}
\label{tab:r2compressed_cloth}
\begin{tabular}{lcc|cc}
\hline\hline
{Method [+ DL Features] \textbackslash Target } & {$Q_{it}$} & {$P_{it}$} & {$\Delta Q_{it}$} & {$\Delta P_{it}$}\\ \hline
{Linear Reg [+5 PCAs]} & 20.02\% & 55.71\% & 2.30\% & 0.27\%\\
{Linear Reg  [+ 5 Similarities]} & 19.23\% & 51.46\% & 2.47\% & 0.31\%\\
{Linear Reg  [+256 Embeddings]}  & 13.28\% & 56.75\% & 0.00\% & 0.00\% \\
{Boosted Trees Reg [+5 PCAs]} & 26.42\% & 63.62\% & 3.55\% & 0.00\%\\
{Boosted Trees Reg [+5 Similarities]} & 25.05\% & 58.53\% & 1.74\% & 0.00\%\\
{Boosted Trees Reg [+256 Embeddings]}  & 26.73\% & 64.67\% & 5.25\% & 0.00\% \\[5pt] \hline \hline
\end{tabular}
\vspace{2mm}
\caption*{ \footnotesize Note: All models are trained on a training set and scores are evaluated on a test set. Predictions use lagged values of time-varying controls. Negative $R^2$ values are set to $0$.}
\end{table}

\pagebreak

\subsection{Homogeneous Elasticity Model}

\begin{table}[h]
\centering
\caption{Estimated price effects based on the partially linear dynamic model (Clothes dataset).}
\label{tab:dynamic_effect_analysis_clothes}
\resizebox{1\textwidth}{!}{
\begin{tabular}{lcccccc}
\hline\hline
  Specification of Control Function (State $S_{t}$)  & {coef} & {std err} & {t} & {P-val.} & {[5.0\%} & {95.0\%]} \\ \hline
{I-1. Linear ($P_{t-1}$, $Q_{t-1}$)}                                  & -0.625 & 0.117 & -5.34  & 0.000 & -0.817 & -0.432 \\
{I-1. Linear ($P_{t-1}$, $Q_{t-1}$, $X^{e}, X^{o}_t$)}                & -0.609 & 0.113 & -5.41  & 0.000 & -0.794 & -0.424 \\
{I-1. Linear ($P_{t-1}$, $Q_{t-1}$, $X^{o}_t$)}                       & -0.566 & 0.115 & -4.92  & 0.000 & -0.756 & -0.377 \\
{I-2. Linear with Interactions ($P_{t-1}$, $Q_{t-1}$, $X^{e}, X^{o}_t$)} & -0.555 & 0.111 & -4.99  & 0.000 & -0.739 & -0.372 \\
{I-2. Linear with Interactions ($P_{t-1}$, $Q_{t-1}$, $X^{sim}, X^{o}_t$)} & -0.561 & 0.112 & -5.01  & 0.000 & -0.745 & -0.377 \\
\hline\hline
\end{tabular}}
\caption*{\footnotesize Note: Standard errors are clustered at the product level. The LR models are estimated using OLS. The PLR model is estimated using DML with cross-fitted boosted trees (sample size: 13,734).}
\end{table}

\break
´
\subsection{Heterogeneous Elasticity Model}

\begin{table}[h]
\small
\centering
\caption{Inference on the price effect modifiers (Model II-3) with embedding features (PLR with boosted trees, Clothes dataset)}
\label{tab:dynamic_effect_analysis_cate_clothes}
\begin{tabular}{lcccccc}
\hline\hline
Modifier & {coef} & {std err} & {t} & {P-val.} & {[5.0\%} & {95.0\%]} \\ \hline
Intercept & -0.681 & 0.141 & -4.832 & 0.000 & -0.913 & -0.449 \\
Lagged Quantity, $Q_{t-1}$ & 0.131 & 0.214 & 0.612 & 0.541 & -0.221 & 0.483 \\
Lagged Price, $P_{t-1}$ & -0.097 & 0.148 & -0.654 & 0.513 & -0.341 & 0.147 \\
Cluster Similarity 0 & -5.983 & 6.046 & -0.990 & 0.322 & -15.928 & 3.962 \\
Cluster Similarity 1 & -11.806 & 15.673 & -0.753 & 0.451 & -37.586 & 13.974 \\
Cluster Similarity 2 & 20.210 & 7.474 & 2.704 & 0.007 & 7.916 & 32.504 \\
Cluster Similarity 3 & 26.046 & 14.188 & 1.836 & 0.066 & 2.708 & 49.383 \\
Cluster Similarity 4 & -28.047 & 11.249 & -2.493 & 0.013 & -46.550 & -9.544 \\
\hline\hline
\end{tabular}
\end{table}

\begin{figure}[ht]
    \centering
    \includegraphics[width=0.8\linewidth]{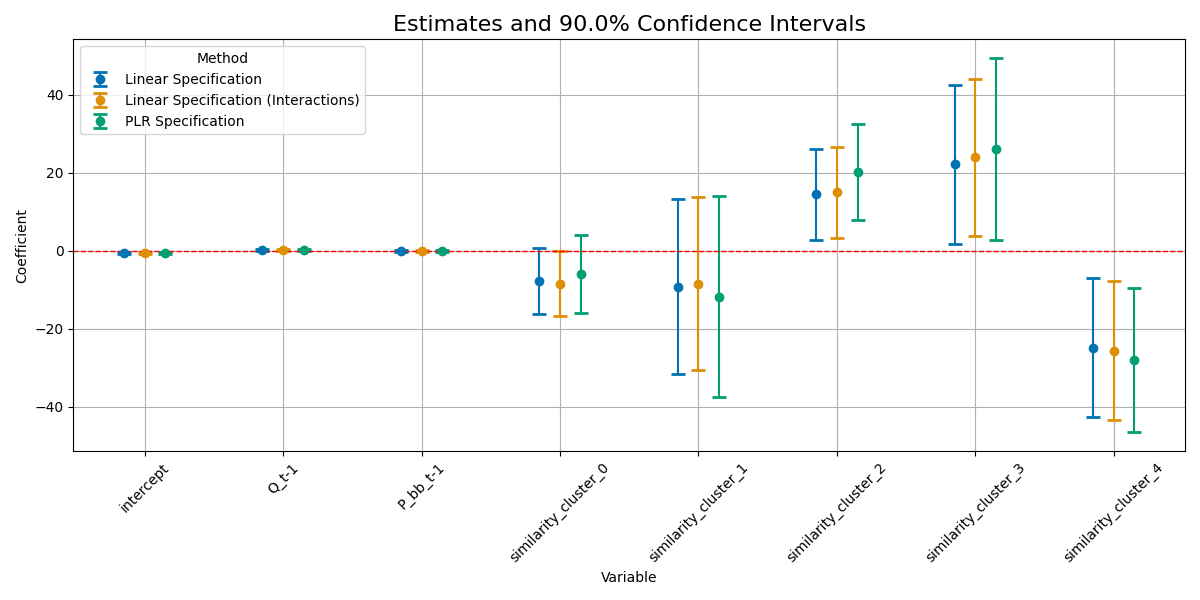}
    \caption{Parameters of the Estimated Elasticity Function and 90\% Confidence Intervals}
    \label{fig:cate_sub_clothes}
\end{figure}

\begin{table}[ht]\footnotesize
    \centering
    \caption{$p$-values for $\chi^2$-test of joint significance of price effect modifiers (Clothes dataset)}
    \label{tab:cluster_significance_dynamic_clothes}
    \begin{tabular}{@{}lcc@{}}
        \hline\hline
        Model & All Modifiers & Similarities Only\\ \midrule
        II.1 Linear Specification & 0.0000 & 0.0019 \\
        II.2 Linear Specification (Interactions) & 0.0000 & 0.0018 \\
        II.3 PLR Specification (Boosted Trees) & 0.0000 & 0.0013 \\ \hline\hline
    \end{tabular}
    \caption*{\footnotesize Note: Standard errors are clustered at the product level. LR is estimated using OLS. The PLR model is estimated using DML with cross-fitted boosted trees (sample size: 13,734).}
\end{table}

\begin{figure}[ht]
    \centering
    \includegraphics[width=0.8\linewidth]{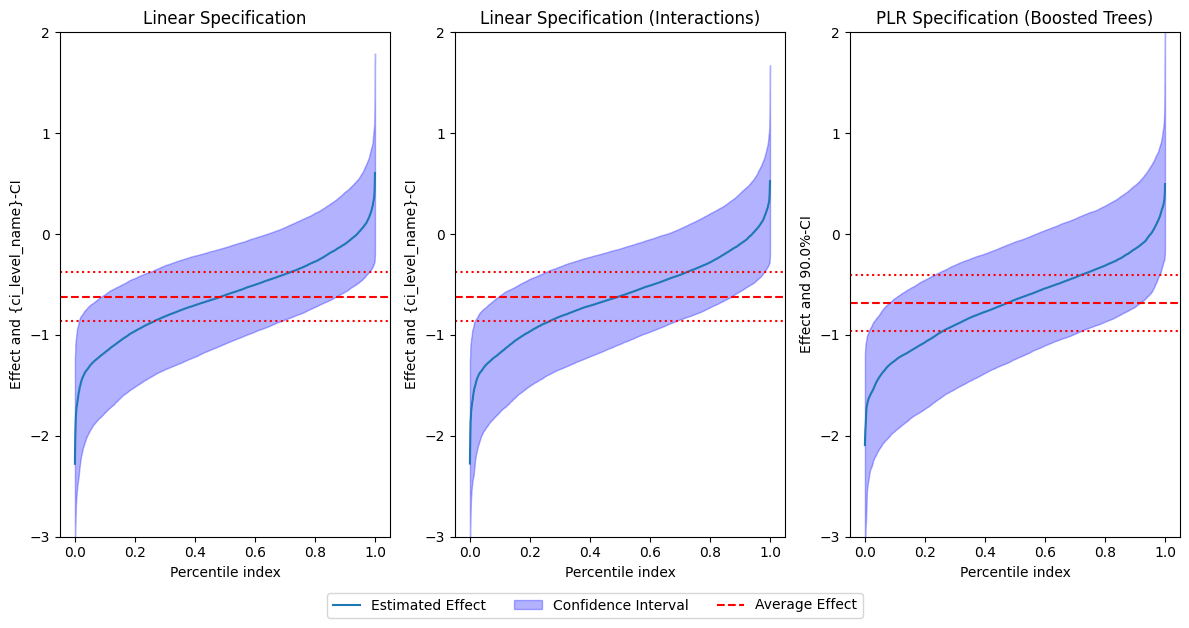}
    \caption{Sorted Elasticity as a Function of Effect Modifiers and 90\% Pointwise Confidence Bands (Clothes dataset)}
    \label{fig:sorted_sub_cloth}
\end{figure}